# Challenges in Blockchain as a Solution for IoT Ecosystem's Threats and Access Control: A Survey

Suranjeet Chowdhury Avik, Sujit Biswas, Md Atiqur Rahaman Ahad , Zohaib Latif, Abdullah Alghamdi, Hamad Abosaq , Anupam Kumar Bairagi

*Abstract*—The Internet of Things (IoT) is increasingly influencing and transforming various aspects of our daily lives. Contrary to popular belief, it raises security and privacy issues as it is used to collect data from consumers or automated systems. Numerous articles are published that discuss issues like centralised control systems and potential alternatives like integration with blockchain. Although a few recent surveys focused on the challenges and solutions facing the IoT ecosystem, most of them did not concentrate on the threats, difficulties, or blockchain-based solutions. Additionally, none of them focused on blockchain and IoT integration challenges and attacks. In the context of the IoT ecosystem, overall security measures are very important to understand the overall challenges. This article summarises difficulties that have been outlined in numerous recent articles and articulates various attacks and security challenges in a variety of approaches, including blockchain-based solutions and so on. More clearly, this contribution consolidates threats, access control issues, and remedies in brief. In addition, this research has listed some attacks on public blockchain protocols with some real-life examples that can guide researchers in taking preventive measures for IoT use cases. Finally, a future research direction concludes the research gaps by analysing contemporary research contributions.

## I. Introduction

THE Internet of Things (IoT) refers to electronic devices that have many built-in sensors and are connected to the Internet. IoT leverages a wide range of applications with integrated support of various communication technologies, protocols, sensor networks, proximity technologies, etc. [1]. According to IoT analytics, the enterprise IoT market grew by 22.4% in 2021 to reach USD 157.9 billion [2]. Figure 1 displays a bar graph showing the number of IoT devices that are worldwide connected from 2019 to 2023 along with predictions for 2030[3]. Further, from 2019 to 2025, it provides an overview of consumer IoT (CIoT), which is primarily responsible for transporting data from the end user[4]. By 2030, there will be 9.7 billion more IoT devices in operation globally than there were in 2020, totaling more than 29 billion. This growth also has a huge impact on the world economy. IoT Analytics projects a 22.0% compound annual growth rate (CAGR) for the IoT market size between 2022 and 2027, bringing it to 525 USD billion [5]. Application of these IoT devices directly influences end-users lives by facilitating the development of smarter home integration, innovating agriculture, building smarter cities, enhancing supply chain management, improving

Corresponding Author: Sujit Biswas

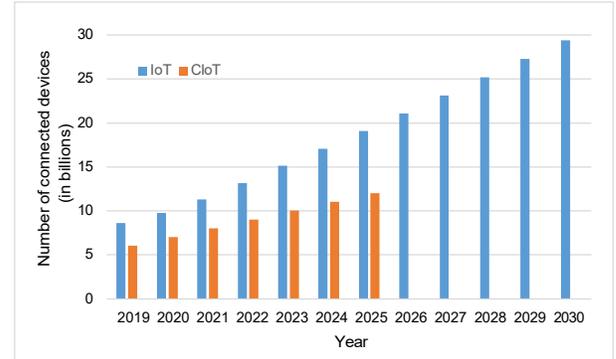

Fig. 1. Generic IoT and consumer IoT (2019-25) usage statistics and prediction

health care [6], consolidating internet-linked manufacturing facilities, reshaping hospitality, etc. For example, in smart home lighting, gardening, security, air quality, water quality, voice assistants, switches, locks, electricity, water meters, and many more private communications [7]. Animal tracking, known as "Intelia" in the agricultural industry, is used for precision agriculture and farming. The convenience of city life, as well as urban security and safety. Another important application is e-healthcare, in which patients are remotely monitored, the elderly are cared for remotely, and so on [8].

### A. IoT ecosystem challenges

Most IoT-leveraged smart systems are controlled from a centralised server which stored the most important information in a centralised location. The information includes industrial production, personal information, and very private information about the users, depending on the use cases. Data leakage from IoT or centralised control systems ultimately discloses the end-user information or company is secret [9]. Existing data can also be contaminated in the same way. The effects of security risks in the context of the IoT can be far more severe than the dangers themselves. This is especially clear in the area of the IIoT, where attacks in the past have caused effects to spread. The use of IoT devices in healthcare for remote monitoring of vital signs has already proven useful during the epidemic [10]. As such, attacks on such devices might compromise patients' privacy and possibly put their lives in peril.

Security and privacy standards must be enforced to prevent any system vulnerabilities or dangers because of the rapid expansion of the IoT [11]. Reliability, scalability, and power consumption are all essential considerations when it comes to the industrial IoT (IIoT) [12] [13]. As a result, typical security measures may not always work. Also crucial in an IIoT context is one or more of these assaults. In an IIoT setting, certain IoT-related attacks can have much more destructive consequences. Layered-based attacks on IoT devices are increasing day by day. In order to fix these device-level security problems, the industry has already come up with a lot of security concerns. Many mitigation techniques have already been implemented in the past. Blockchain-based security methods, however, are excellent for protecting the Internet of Things.

Since blockchain makes IoT secure and decentralised, it is a good solution for the aforementioned issues. A blockchain is essentially a series of blocks, each of which is timestamped and connected to the others by cryptographic hashes[14]. It functions as a kind of decentralised ledger, with its data being shared among a network of users. Furthermore, IoT is a promising field in which many smart applications are being created. IoT applications use sensors, smart objects, and actuators. Many people are interested in studying the usage of blockchain technology in various industries because of its transparency and integrity qualities, which cannot be changed retrospectively without altering all the following blocks. To support seamless sign-on, blockchain can be used to securely track and manage digital identities. Digital signatures based on public-key cryptography serve as the foundation for most blockchain-based authentication solutions. Each transaction must be signed by an authorised private key in order to identify a user's identity on a blockchain. There are many uses of blockchain in the IIoT, but there are also many ongoing obstacles to overcome. To eliminate the requirement for a trusted third party intermediary, blockchain technology is one example[15]. While the potential of this technology is undeniable, several obstacles must be overcome before it can be employed in a meaningful way. When using blockchain technology to an IIoT application,it is important to take into account a broader set of concerns throughout the solution's design phase[16].

*B. Blockchain in IoT eco-system*

Security and privacy must be maintained to prevent system vulnerabilities and threats brought on by the expanding IoT [11]. When it comes to the IIoT, reliability, scalability, and power consumption are all crucial factors to take into account. Therefore, standard security measures might not always be effective. One or more of these challenges are also essential in the context of the IIoT [12]. Certain IoT-related attacks may have far greater detrimental effects in an IIoT environment.

As blockchain is secure and decentralised, it can be an excellent solution to the issues[17]. Blockchain is essentially a collection of timestamped blocks connected by cryptographic hashes [14]. It functions as a decentralised ledger, with its data being shared among a network of users. Every transaction is recorded in immutable blockchain consiquently the blocks are added into ledgers which is also a breakable chain. Immutability in the context of IoT systems can stop data manipulation. Blockchain technology (BCT) has made it possible to handle distributed transactions, the building blocks of the IoT ecosystem, in a creative and exciting way [18]. The core security mechanism of BCT, which was initially used to defend a single consortium as a coin, is currently utilised in both public and private applications. The elimination of centralization and the automated management of a secure flow of real-time data across IoT devices are also primary drivers for Blockchain's acceptance in the IoT. People can control who sees their data and under what conditions thanks to Blockchain technology. Because the public ledger can be traced back to the point at which changes or data tampering were necessary to commit a new block, the Blockchain can be an excellent source of trustworthy traces and artefacts. However, Blockchain has its limitations, much like any emerging technologies, and using it in crucial IoT networks is a very risky proposition. For instance, there are still many unanswered problems regarding the use of IoT sensors to supply chain management and the state of data security in the healthcare industry.

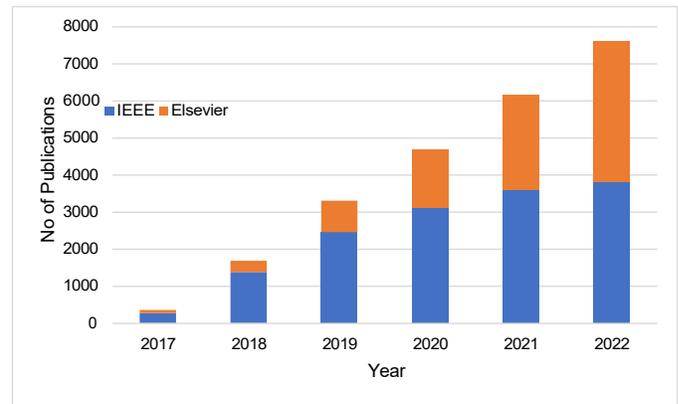

Fig. 2. Blockchain contributions published from 2017-22 in IEEE and Elsevier (Science Direct)

*C. Current status*

We combed through publications from the most prestigious academic journals and publishers like IEEE and Elsevier to get a sense of Blockchain and IoT's overall impact in the academic world. As shown in Figure 2, there were 24500 related articles on Blockchain published in IEEE and Elsevier only from 2017 to 2022. This huge number of articles proves the importance of these technologies in various applications. The majority of these contributions recommend BCT for IoT system security while they taking a lot of fine-grained aspects into account, including thinking about security, including data security, authentication, multi-user access, and so on. For example, Sookhak et al. [19] summarise blockchain-based IoT access control for healthcare systems. It describes different types of key-based encryption for healthcare security. Ghaffari et al. [20] discuss access control for integration in business databases. It claims that blockchain and other distributed ledger technologies are having an effect on how businesses operate today. Bagga et al. [21] summarise device-to-device access management using blockchain technology. Pal et al. [22] discuss access control mechanisms in the IoT that are developed on three commonly used access control mechanisms:

role-based access control (RBAC); attribute-based access control (ABAC); and capability-based access control (CapBAC). Shakarami et al. [23] discuss the blockchain-based administration of access in smart home IoT. Smart contract security and device-cloud communication are discussed. Sekaran et al. [24] discusses IoT automation through mobile edge computing provided by 6G and the blockchain. Traditional areas of interest for a range of purposes were represented, such as intelligent manufacturing, supply chain management, the food sector, the intelligent grid, healthcare, media and digital rights administration, agriculture, the internet of automobiles, and unmanned aerial vehicles. Abubakar et al. [25] briefly describe the process of creating a distributed system of authentication and access control for medical data acquired by wearable sensors. In [26], authors discussed the growth of decentralised end-to-end manufacturing applications. Smart contracts are agreements between people who want to use a service and the people who make the service (DApps) on demand. They built a decentralised system in which the nodes checked and monitored each other. This makes for a safe marketplace where data resources can be traded. More details are covered in related work sections. However, in earlier reviews, the authors only addressed one site, either through threat detection [27], access control [28], new decentralisation techniques [29], or architectural limitations[30]. They particularly focused on a particular aspect of access control and other blockchain-based strategies for thwarting threats. On the other hand, we have taken into account every area of security, including the threats that exist in IoT ecosystems, the difficulties these threats provide, and potential solutions. We concluded from our literature review that access control systems can address the majority of security concerns. We have therefore focused on access control problems and blockchain-based solutions. Additionally, we have outlined current blockchain concerns and the state-of-the-art in recent cyber-hacking, with justifications. A study on attacks on blockchain will guide researchers to improve IoT security because of the breadth and expressiveness of the information regarding those threat mitigation approaches. This article groups the key security concerns of the IoT and dig into the research gap required to focus on making this domain more conducive to overcoming real-life problems. In sharp contrast to the survey articles found in the literature, this research contributes:

- A list of issues with details categories that may result from different attacks on IoT systems.
- A high-level overview of access control problems and typical issues.
- A collection of illustrative blockchain-based solutions that have recently been offered in articles.
- A list of probable attacks on blockchain systems along with historical illustrations.
- A useful research direction based on our study.

The rest of the paper is organised into another six sections. Section II describes the previous related reviews and introduces the reason for this review. IoT system threats have been presented in Section III. We have categorised various threats in this section. Access control challenges arise due to various threats and their solutions as well as blockchain-based remedies have been discussed in Section IV-B. A historical attack on the blockchain protocol with technical details has been presented in Section V. Finally, we have presented a brief summary of future research directions and concluded the overall contribution in Section VI and VII respectively.

## II. RELATED WORK

As IoT devices carry consumer information, this raises security and privacy concerns, especially for end-users' private information. Moreover, when the devices are used in smart industries, it also raises some issues. Because of resource constraints, adopted technology is vulnerable. Malicious campaigns that flood targets with millions of requests are called DDoS assaults, and they can affect a large number of IoT devices. As there are currently no authentication standards for IoT devices, they pose a risk to vital infrastructure [31]. The constant connectivity of IoT devices also leaves them open to cyber threats. Overall, IoT applications need to be scalable and reliable. Blockchain technology can make sure that the device's readings are correct and prevent bad information from getting out. Blockchain not only provides data security but also allows for individual devices to be tracked and recognised. This is a must if you want to keep devices that connect to the internet safe. Considering the cyber-security properties of blockchain, a huge number of researchers suggest using blockchain to overcome security and privacy issues in the IoT ecosystem which has been reflected in Figure 2. To understand the current implementation state of BCT and IoT system integration, we have searched through 3798 of connected research publications using advance searches and related keywords. A thematic graphic that highlights the value of BCT and IoT integration may be found in Figure 3. After multiple rounds of filtering and removing duplicates, we finally worked with 3695 articles after screening a small number of publications based on titles and abstracts. Finally, we assessed 195 articles are considered in this our review process beyond some various online sources (e.g., white papers, blogs, etc).

Moreover, we have screened this section summarises a few of the review papers that conclude the existing contributions to overcoming the IoT ecosystem challenges. The number of prior review articles from 2017–2022, published in IEEE and Elsevier, is shown in 4. This contribution differs greatly from state-of-the-art contributions, even if they are connected in a broader context and touch on themes like access control, Blockchain integration with the IoT, and Blockchain. Finally, the main points of departure from previous appraisals have been discussed.

Mohanta et al. [32] discussed Artificial Intelligence (AI) and Blockchain integration with IoT, where they focused on private key encryption, an important idea that could be used in many places, like the medical field, the IoT, manufacturing, and supply chain management, all of which are also talked about in the context of AI-IoT integration. How blockchain can ensure IoT devices' security, especially IoT-generated data security, has been well analysed in almost every application domain. The number of review articles well summarises those research contributions from various points of view. For example, the effects of having a lot of IoT devices, slow computers, not enough bandwidth, and bad radio connections on how well blockchain works were looked at. Numerous DAG and GHOST protocol use cases are also discussed for securing the communication data in [33]. Again, Memon et al. [34] discuss hybrid-IoT as a three-tiered architecture: fog, edge, and cloud. Hybrid IoT uses three different communication architectures that work well for different use cases. Hybrid IoT gets around the problems with both existing infrastructures (e.g., cloud-based IoT and Blockchain-based IoT) and meets the standards expected of future IoT. Gao et al. [35] analysed Blockchain technology from the vantage points of the data layer, the network layer, and the application layer, paying special attention to the activities or implementations of each. The author evaluated the literature on a wide range of use cases and found that Blockchain's permanent record is well suited to the decentralised architectures of the IoT, the cloud, and even the edge computing environments known as fog and edge computing. Furthermore, composite layered access control methods based on Blockchain, as well as the opportunities and designs of the Blockchain of Things (BCoT) are discussed in [36]. Uddin et al. [37] Although in a broader context those review articles are related and touch on topics including access control, Blockchain integration with the internet of things, and Blockchain, this contribution is significantly different from state-of-the-art contributions. Finally, we have mentioned the key differences between this contribution and existing reviews. By using Blockchain technology, healthcare can be changed and improved in ways like interoperability, which means that patients can get access to their medical records and medicines, and hospital assets can be tracked safely throughout their entire life cycle.

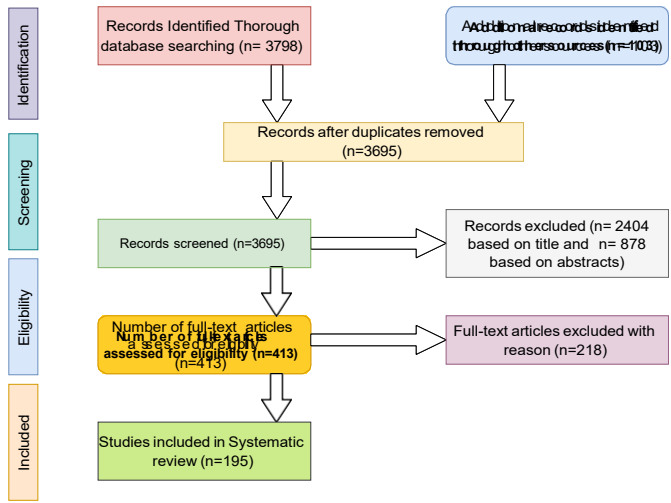

Fig. 3. Brief but comprehensive summary of the research's methodology and papers selection processes; main findings and conclusions drawn from the meta-analysis and synthesis

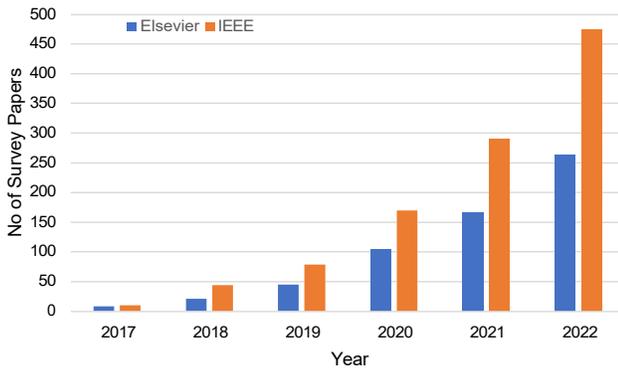

Fig. 4. Recent review articles somehow related to blockchain

As blockchain is a well-known cybersecurity technology, a large number of contributions have recommended blockchain as a cyber-security tool such as access control, transaction security, etc. in the IoT. For example, Feng et al. [42] discussed access control methods for different verification circumstances such as cryptographic security, distributed access control, synchronised databases, etc. Namane et. al. [48] IoT applications are also used to categorise recent Blockchain-based access control frameworks. It is feasible to pinpoint the results of each domain by analysing these efforts according to the domain of application. Similarly, by using a local private blockchain, a lightweight security mechanism can be implemented to satisfy IoT needs, as discussed in [49]. Without relying on third parties, secure communication with many entities is ensured through multiple-chain code-based access control (TTP). In addition, Da et al. [47] discussed the access control-related contributions where device-level authentication was privileged. It has been talked about how Blockchain, which is a platform for a distributed and transparent database, could improve IoT security for public services. The Ethereum blockchain, data management, and authentication techniques discussed by Panarello et al., in [39]. Device manipulation and how it affects data management as a solution for an open market are also talked about. Key management is solved by securely distributing the secret key using the elliptic curve Diffie-Hellman key exchange protocol. Besides, Han et al. [51] considered the Intel SGX Enclave security zone for processing sensitive data in edge node data processing and transmission. Similarly, Al et al. discuss a layered threat mitigation approach based on cloudlets and blockchain in their [46]. They talked about things like making sure that data is stored and processed in a decentralised way, addressing worries about the safety of anonymity, and making sure that the steps needed for effective authentication in public services are clear. Comparisons of various solutions in the fields of cryptography, consensus mechanisms, network security, and application-level security for the P2P business model are discussed in [43]. Likewise, compare the security of Blockchain to that of a traditional cyber network and examine both in light of numerous cyber-attack use cases in [38]. Blockchain-based threat mitigation techniques and methods are covered in Salimitari et al. [41] where they covered the drawbacks of some blockchain protocols such as Hyperledger Fabric, Sawtooth, IoTA, and Ethereum. They focused on better options for IoT networks and applications in terms of throughput, latency, computational overhead, network overhead, scalability, and privacy.

TABLE I
SUMMARY OF RECENT RELATED CONTRIBUTIONS

| Paper | Published | Access Control | Security Threats | Blockchain | Comments |
|---|---|---|---|---|---|
| [38] | 2018 | X | | X | Cloud Security. |
| [35] | 2018 | X | | X | Cloud security. |
| [39] | 2018 | | X | X | Smart manufacturing. |
| [40] | 2018 | X | | | Gervais technique. |
| [41] | 2019 | X | | | Consensus protocol. |
| [42] | 2019 | | | X | Privacy preservation. |
| [33] | 2019 | | X | | GHOST protocol. |
| [43] | 2019 | | X | X | TCP/IP four-layer Authentication. |
| [32] | 2019 | X | | X | Intelligent transportation system. |
| [36] | 2020 | | X | X | Smart home application. |
| [44] | 2020 | | X | | Open and distributed ledger. |
| [34] | 2020 | X | X | | Distributed consensus algorithms. |
| [45] | 2020 | | | X | Anti-counterfeiting Systems. |
| [46] | 2021 | X | | X | Cloudlet computing. |
| [47] | 2021 | | X | X | Cloudlet-dew architecture. |
| [37] | 2021 | | X | X | Access control mechanism. |
| [48] | 2022 | X | X | | Eliminating Trusted Third Party (TTP). |
| [49] | 2022 | X | X | | Multiple layers for privacy protection. |
| [50] | 2022 | | X | X | IoT in healthcare. |
| [51] | 2022 | X | | | Software guard extensions (SGX). |
| This Paper | 2023 | | | | IoT ecosystem's threats, security challenges, solutions, attacks on blockchain |

The most recent relevant articles are included in Table I along with their major contributions. We reviewed most of the previous related contributions that addressed issues with the IoT ecosystem and blockchain-based solutions. Each review paper skillfully organised the most recent contributions, which individually span AI, computing, communication, access control, and frameworks. Contrary to popular belief, the purpose of this review is to investigate the recent contributions that are focused on threats to the IoT ecosystem, mitigation challenges and solutions, and how blockchain can help overcome these challenges. We have broadly investigated various types of access control to ensure end-user security. To the best of our knowledge, this is the first review that covers every aspect of IoT ecosystem security, from potential threats to solutions.

## III. SECURITY THREATS IN IoT SYSTEMS

One of the most popular ways to steal data is by altering the firmware of an IoT device or, in some other manner, regulating the transactions that the IoT creates. This frequently leads to IoT security problems. Access to financial or personal information is regularly sought in this way. We classified every possible attack into five categories based on the vulnerable zone or targeted area. These include physical attacks, attacks on communications or networks, attacks on software or logic, attacks on data, and attacks involving management. Figure 5 illustrates the potential attacks in every category that have been mentioned in various academic articles. A brief summary of previous relevant contributions is also summarised in the Table II.

### A. Physical attacks

Gadgets are commonly unlocked and accessible to numerous individuals. Lack of physical security puts the IoT ecosystem in danger. An attacker is close enough to a target network or system in a physical attack. For example, *malicious Code* can be injected by an attacker that can repeatedly perform various attacks [52]. Instead of code injection, sometimes intruders try to control or inject a managed node in the overall ecosystem which is used to manipulate data flow among legal nodes in the network [53]. Besides data theft or controlling the network, interruption in communication flow is another common type of physical attack. For example, *radio RF interference or jamming* is a well-known attack in wireless sensor networks because of the DoS attack on RFID tags/sensor nodes [54]. It can stop the overall communication system in the WSN (e.g., vehicular network). In addition due to *permanent denial of service (PDoS)* IoT devices can be sabotaged and rendered inoperable by plashing, a type of DoS attack. To carry out the attack, a virus is used to delete the kernel or upload a corrupt BIOS [55]. By *Side Channel attack* an attackers can obtain encryption keys by exploiting flaws in the system's hardware, such as timing, power, and other errors [56].

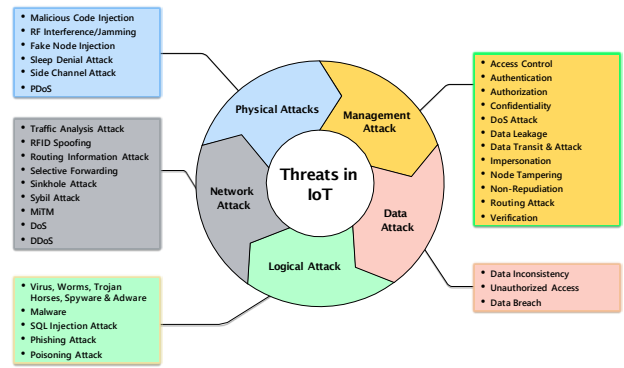

Fig. 5. Every possible attack in various broader attack categories in the IoT ecosystem

### B. Network attacks

Network attacks are unauthorised intrusions into a company's computer network or its associated digital resources.

Hackers target networks in order to alter, delete, or steal sensitive information. Perimeters are a common target for hackers seeking entry into a network. IoT network systems can be manipulated to cause harm through network assaults. It can be close to a network to get started. As most IoT devices are resource-constrained and don't have strong authentication mechanisms, intruders use communication mediums to eavesdrop on the system. For example, *RFID Spoofing* where an RFID signal is spouted by the attacker to obtain access to the data on the RFID tag. Hackers can then use the source tag ID to transmit data that appears to be real [57]. Due to *routing information attacks*, routing information is altered and causes problems in generating routing loops, issuing error messages, or causing traffic to be routed in the wrong direction [58]. Similarly, an attacker can get network information by *Traffic Analysis attack* even if they are not physically adjacent to the network by eavesdropping on the communications between devices [59]. Losing control of the network can allow *selective forwarding attack* in which the rogue node might change, drop, or deliberately forward some messages to other nodes in the network. Because of this, the information that reaches its final destination is insufficiently [60].

Bandwidth stability is very important for active connectivity. Lower bandwidth can generate *Wormhole Attack* which can compromise tunnel packets between two points [61]. For any reason, a malicious node can occur, resulting *Sybil Attack* which is introduced as Sybil nodes can change their identities and travel throughout the network, resulting in serious resource misuse [62]. In an IoT environment, malicious people who can assume many identities and operate a distributed network can take control of the governing network. One of the most common and well-known attacks in communication technology, especially wireless communication, is *Man-in-the-Middle Attack (MiTM)* which allows an attacker to listen in on or watch how two IoT devices talk to each other to get to their private data. Wearable IoT devices are controlled through smartphone applications that can cause MiTM to compromise application permissions [63], [64].

Although information theft is the main goal of most attacks, the most perilous type is the *Denial-of-Service (DoS) Attack*, which renders a system or network completely inoperable[65]. This is because a denial of service attack might overwhelm the target with traffic or supply data that causes it to fail. When the server or other Internet of Things devices are inaccessible due to a denial of service attack, the entire network is at danger. In addition, in *Distributed Denial of Service (DDoS) Attacks*, several systems coordinate to flood a single host with malicious data. For example, an adaptable IoT system can be targeted for control by a group of users [66].

### C. Logical attacks

Logical attacks aim to interrupt the logical flow of a device's actions, which can be accomplished in one of two ways: either by physically getting access to the device in question and then launching an attack on it, or by leveraging services and trafficking hacks. Attacks against an IoT system's software or security weaknesses can be classified as software attacks or logical attacks[67]. An attacker can get access to sensitive information, steal data, or launch a DoS attack by infecting a system with a virus, worm, Trojan horse, spyware, or adware[68]. Malware on IoT devices may theoretically propagate to cloud services and data centres. It's also possible for malware to attack IoT devices and then spread to other cloud services or data centres[68].

Often attackers employ SQL injection to exploit network flaws and obtain access to database administration. SQL injection occurs when a hacker intentionally enters malicious code into a web form in order to steal sensitive data. [69]. The database or website will execute the malicious commands and give the attacker access to the data they requested if it is not properly protected. In IoT system, Phishing is very common attack due to lack of technical consciousness and carefulness of end-users. It is an online social engineering technique intended to deceive users into divulging sensitive information or downloading malware-spreading applications. Emails containing harmful attachments or links to fraudulent websites are the most typical method of phishing attack [70]. Again, domain names are converted into IP addresses via the DNS, or domain name system. An attacker must first get access to one of the DNS servers in order to modify a specific domain's or website's IP address so that users trying to access it will be redirected [71]. DNS Poisoning is the process by which an attacker creates a fake DNS server response and delivers it to a DNS server to tamper with data that was already cached there.

### D. Data attacks

As the IoT develops, more and more cloud storage is being put to use. With cloud computing, it's simple to deploy virtual servers, start databases, and construct data pipelines for IoT applications[72]. The cloud can aid in improving data security by providing suitable authentication mechanisms, firmware and software update systems, etc. In this section [73], we look at the most typical forms of data theft in the IoT. A data breach or memory leak is the unauthorised disclosure of personal, sensitive, or confidential information[74]. A "Data Inconsistency" occurs whenever there is a discrepancy between the data in motion and the data stored in a centralised database as a result of a data integrity attack[75]. In addition, access control refers to the steps taken to allow only approved users access and deny access to unauthorised users[76]. Unauthorised access gives bad actors a chance to steal private information or claim ownership of data.

### E. Management attacks

The growing use of IoT devices is changing the landscape of the enterprise network. The IoT has the potential to greatly enhance operational transparency and productivity. Networking these IoT devices allows for streamlined and centralised management across all locations. But, there are serious security concerns for everyone involved when it comes to IoT devices. A compromised IoT device can be exploited by a hacker for network lateral movement or as a bot in bigger malware campaigns to launch different attacks on internet-connected devices. With the high number of diverse low-end devices linked and transferring and processing real-time data, the IoT poses enormous issues in terms of access control [87]. Distributed design, an adhoc networking model, and many sensors all contribute to bandwidth issues between IoT devices. Authentication methods now in use include passwords, smart cards, and fingerprints, although none of these methods operate on low-end devices. Nodes in an IoT application must be authorised to ensure safe data access or transmission [88]. Data must be authorised by a centralised authority or decentralised network before it can be exchanged between devices [89]. The network must be given permission before any data can be processed. In a decentralised IoT design, determining how to authenticate each node in the network is a major difficulty. Again, Fog nodes, or edge nodes, are utilised to process real-time data in an IoT network since they have large numbers of sensors [90].

TABLE II
A SUMMARY OF CONTRIBUTIONS CONSIDERED VARIOUS ATTACKS

| Paper | Attack types | | | | |
|---|---|---|---|---|---|
| | Physical | Network | Logical | Data | Management |
| [77] | | | | X | X |
| [78] | X | | X | X | |
| [79] | X | | | X | X |
| [80] | X | X | X | | |
| [81] | X | X | | | X |
| [82] | | X | X | X | |
| [83] | | X | | X | |
| [84] | X | X | X | | |
| [85] | X | | | X | |
| [86] | X | | | X | X |
| This paper | X | | | | |

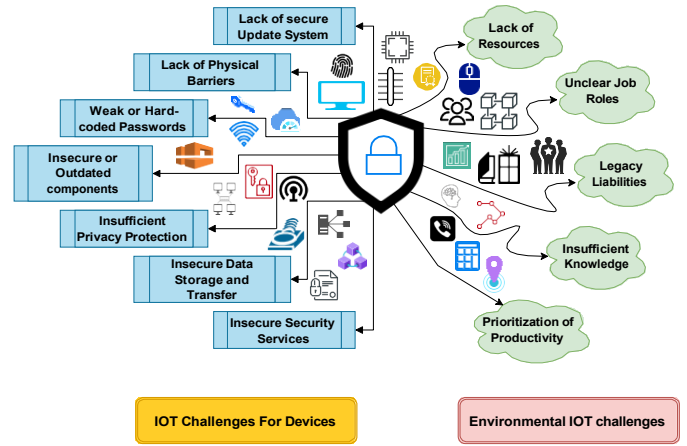

Fig. 6. Access control challenges and it's affected regions in IoT ecosystem

Maintaining the privacy of each node is essential. If the nodes are not secure, important data could be hacked. DoS attacks can take advantage of the CIA Triads' vulnerability in the "availability" area, rendering the server inaccessible to its users[91]. By flooding the server with requests, the attacker hopes to steal data and slow down the connection. Data leakage is the unintentional disclosure of confidential information, either electronically or physically, outside of a company[92]. Intellectual property, medical records, confidential information, consumer data, etc. are all examples of the types of information that might be compromised by a leak. This kind of assault targets the channel through which messages are transmitted.

Man-in-the-middle (MITM) and sniffing are the most popular methods of data eavesdropping [93]. To get access to sensitive information, an attacker may try to impersonate a trusted individual. It is possible for an attacker in this type of attack to gain access to a victim's database by inserting malicious code into a programme running on the victim's machine [94]. Again, The IoT application's device efficiency and reliability are difficult to achieve. Verification in an IoT system can be difficult due to the difficulty of updating hardware components [95]. An assault known as "node tampering" aims to physically alter nodes by physically replacing the complete node or a portion of the node's hardware [96]. During the forwarding process, the attacker places malicious nodes on the network, which affects the routing path [97]. Also, Any transaction or operation in a system that is not authenticated is called non-Repudiation. When it comes to an IoT system, there is a lot of sensitive information that must be authenticated across different nodes [98].

*F. Potential solutions*

There are several issues with the classic client-server IoT setup; access-control issues are considered the reason behind most of these attacks. Recent contributions have provided diverse solutions in various ways . For example, DDoS attacks are a type of cyberattack in which the target is inundated with so much network traffic that it is unable to handle it normally [99]. While conventional IoT implementations tend to be centralised [100], Blockchain's decentralised security design ensures that each device in a network is protected in its own right [101]. Threats like sleep denial attacks, side-channel attacks, and others are mitigated by using Blockchain [18]. The effect on data security is another concern with integrating the IoT and Blockchain [102]. Security for TLS-based protocols like Message Queuing Telemetry Transport (MQTT) [103] and Constrained Application Protocol(CoAP)[104] is dependent on a governing authority and key infrastructure [105]. Alternatively, in Blockchain networks, each device could have its own globally unique identifier (GUID) [106] and asymmetric key pair [107]. This would reduce the time required to analyse the data and make it more difficult for hackers to gain access. Blockchain's ability to track and record data from all devices on a network means that enterprises can more easily detect and prevent security threats, including MiTM attacks, Sybil attacks, replay attacks, and more [108].

IV. ACCESS CONTROL CHALLENGES AND SOLUTIONS

IoT devices can't fully protect themselves because of resource limitations, which makes it very challenging to offer secure authentication. However, a number of academic contributions offer a list of common problems and solutions. This section presents an overview of many contemporary problems and recent contributors' blockchain-based solutions. Figure 6 illustrates several access control issues in IoT ecosystem. It shows the concern that relate to access control challenges in the IoT ecosystem.

*A. Access control challenges*

As the majority of IoT devices have limited resources, including battery life, network bandwidth, and computing power, there are many difficulties in setting up authentication and access management in an IoT environment. The majority of traditional authentication methods require a significant amount of communication overhead, making them impractical for usage with the constrained capabilities of IoT devices [109]. For instance, in 2019, at least 84 distinct authentication algorithms [110] were either suggested or implemented for use in IoT contexts, according to the work of several researchers. Ensuring the security of devices and networks is complicated by the absence of standards and IoT-specific access control methods. Although the increased use of Internet of Things (IoT) devices and various applications has greatly improved our daily lives, it has also brought up several difficulties, particularly in the area of access control.

- **Device Heterogeneity:** IoT devices, such as sensors, wearables, appliances, and industrial machinery, are proliferating quickly and come in a wide range of sizes and designs [111]. They used a range of communication functions and protocols. In an ecosystem, it can be difficult to maintain uniform access control for all devices, even though each one needs procedures for authentication and permission. A varied assortment of IoT gadgets by putting into practise an AoT prototype with various levels of security, utilising a range of platforms with various levels of computational capacity is presented in [112].
- **Resource Constrain:** IoT devices, especially consumer IoT, have a limited amount of energy, memory, and processing resources. It cannot be installed with highly computational access control software or encryption technology, which opens many doors to security threats.
- **Scalability:** An IoT ecosystem's volume extends daily for its' fine-grained automation services. In a large ecosystem (e.g., a smart city), thousands or even millions of connected devices can be found in IoT systems. The heterogeneous nature of IoT environments makes it more complex to adopt traditional centralised access control, and continuously extensible networks make it difficult to manage access control policies and ensure scalability [113].
- **Interoperability:** Interoperability is a prerequisite for a smart automation system and is only achievable through data sharing and communication between the components, therefore they must cooperate across platforms, suppliers, and communication protocols [114]. Due to the heterogeneity of IoT devices, achieving seamless interoperability while retaining safe access control becomes difficult [4]. DAAC [18] and Drupal [115] are such framework that expands access control functionality for collaborative environments.
- **Security and Privacy:** End-user data is gathered and processed by consumer IoT; the majority of this data is private and sensitive, particularly for wearable IoT. Strong access control methods, encryption, secure authentication, and authorisation protocols are necessary to ensure data privacy and security. IoT devices are also prone to security flaws, and controlling access across a large number of potentially weak devices is difficult [116].
- **Smart Automation:** It is apparent that the IoT will see frequent additions and removals due to technological developments that make the network dynamic. Maintaining access control policies, revoking access for departed devices, and ensuring ongoing oversight and enforcement of access control rules are all very difficult tasks.

IoT security is a technique that keeps IoT systems secure, and IoT security solutions may help patch vulnerabilities as well as defend against threats and breaches, and discover and monitor hazards. The majority of these breaches happen as a result of weaknesses in access control regulations. Many recent contributions have offered a variety of technological approaches to overcome these common difficulties, which are illustrated in Section IV-B.

*B. Typical Solutions*

A few potential tactics include the development of standardised access control frameworks for IoT devices, the use of basic authentication and authorization protocols, the use of cloud-based access control services, the adoption of encryption and secure communication protocols, and regular firmware updates to address security flaws. Enterprises should prioritise security and privacy in their IoT installations and conduct regular security assessments and audits in order to identify and minimise access control risks. Many recent research contributions suggest various access control policies to overcome the existing challenges. Table III stated a few related contributions and their covering issues, comparing them with this contribution.

In **Attribute Based Access Control(ABAC)**, network attributes (i.e., user identification, device kind, location, time, and other contextual data, etc.) are used to access control. ABAC is a flexible access control model that makes access control choices by taking into account a variety of user, device, and environmental aspects. It also enables dynamic decision-making depending on shifting circumstances. Ciphertext-Policy attribute-based Encryption (CP-ABE) [130] is such an ABAC that avoids data tampering and eliminates a single point of failure with blockchain technology, where smart contracts improve the IoT by reducing authentication. Similarly, authors in [131] illustrate that the IoT, IoT devices, gateways, and attribute servers are controlled through blockchain. Access control and requirements also vary based on responsibility or roles. **Role-based access control (RBAC)** is a model that assigns permissions based on preset roles, where each user or device is assigned a specific role with associated privileges, and access. Authors in [132] proposed an architecture for ensuring role-based access control of web-enabled things in a secure and scalable manner using different encryption technologies. RBAC provides a structured and scalable approach to managing access control within IoT systems. In technology, trust is ensured by a satisfactory performance rating, which in some applications helps to fix access control. **Trust-based access control** is also a well-utilised method in various use cases, while trust comes from the reputation of performance [133]. In the IoT ecosystem, trust-based access control focuses on building relationships of trust between entities. Based on variables including device reputation, previous behaviour, and authentication measures, trust can be assessed. Depending on the level of confidence a given entity enjoys, access decisions are made. Protection against unauthorized access and other security risks, such as those posed by a multi-tenant cloud platform, denial-of-service attacks, insecure interfaces and APIs, malicious attackers, cloud service abuse, data loss, etc., is provided via trust-based access control [133]. **Context-aware access control** is another kind of approach that takes into account the contextual information surrounding access requests [134]. It considers factors such as location, time, network conditions, and user behaviour to make access decisions. For example, access may be granted if the request comes from a trusted device within a specific location during business hours.In 2021, the cloud-based nature of e-health systems presents potential security risks, including those related to patient data confidentiality and privacy. If malicious users are successful in their cyber-attacks, important patient information may be exposed [135].

TABLE III
A BENCHMARKING OF RECENT REVIEWS WITH OUR CONTRIBUTIONS ON BLOCKCHAIN-BASED ACCESS CONTROL SOLUTIONS

| Reference | ABAC | Fair Access | Distributed key management | Token Based access control | Control chain |
|---|---|---|---|---|---|
| [117] |   | X | X | X | X |
| [118] | X |   | X | X | X |
| [119] | X | X | X | X | X |
| [120] |   | X | X | X |   |
| [121] | X |   |   | X | X |
| [122] | X | X |   | X | X |
| [123] |   | X |   |   | X |
| [124] | X | X |   |   | X |
| [125] | X |   | X | X | X |
| [126] | X | X | X | X |   |
| [127] | X | X | X | X | X |
| [128] | X |   | X | X | X |
| [129] | X | X | X | X | X |
| **Our Work** |   |   |   |   |   |

Moreover, **Fine-grained access control** for granting or denying access at a granular level [136], often based on individual resources or data elements within an IoT system. It allows for precise control over what resources or functionalities can be accessed by specific users or devices. Likely, **Secure device onboarding and provisioning** is also used to ensure secure onboarding and provisioning processes for IoT devices by validating device identity [137], establishing secure communication channels, and securely configuring access credentials during the initial setup phase. Device-to-device communication security is confirmed using any **Secure communication protocols** such as Transport Layer Security (TLS) [138].

IoT devices are increasing daily, which requires a scalable, decentralised, interoperable, and dynamic access control platform. Using centralised, typical access control mechanisms scalability can be achieved for relatively small-scale systems, but it is challenging to handle the vast number of interconnected devices in an IoT ecosystem. Likewise, the heterogeneity of devices makes it more challenging to reach uniform access control mechanisms which are very important to ensure a compatible and interoperable unified platform. In addition, the lack of standardised protocols for access control makes it more complex. Ensuring the privacy of end users using typical access control technologies may not adequately address these concerns, as they primarily focus on device authentication and authorization without considering privacy protection mechanisms.

Addressing these limitations requires the development of novel access control solutions tailored specifically for the IoT ecosystem that should be scalable, distributed, platform-independent, dynamic, and interoperable. Many recent contributions by researchers consider blockchains a secure solution to overcome those challenges, which have been described broadly in Section IV-C.

### C. Blockchain as a Solution

Blockchain has become more popular in recent years because of ideas like decentralized structures, security, and the fact that it can't be changed. For example, Tan et al. [120] described ABAC and control chain method, where the Ethereum private chain-based access control scheme is practical and efficient, allowing for private, secure access in the distributed IoT setting. Likely, ABAC method for flexible and granular control over medical records is discussed in [139], [117], [18]. They were able to store encrypted messages using the hyper ledger blockchain protocol, and with the application of intelligent contracts, they made the messages immutable. Fig.7 illustrates Establishing regulations and techniques for controlling access to IoT devices and networks is essential for maintaining security. Due to the decentralisation, immutability of ledgers, and trust-less capacity of blockchain, many recent contributions introduce it as a cyber-security tool, especially for access control.

A control chain method for securing access control is discussed in [126] that decide on which peers to interact and share data, blockchain nodes employ the random neighbor selection (RNS) method. The control chain technique is used to construct optimization frameworks at the consensus, data, network, or a mix of these layers [140]. Consensus algorithms targeting access control also proposed in many recent research works such as Patil et al. [118] included new consensus algorithms for securing transactions that ensures fair access control method named Ripple Protocol Consensus Algorithm (RSPCA). Blockchain based distributed access control methods is proposed in [129] with a very descriptive manner. Another distributed access control-related methods proposed in [123] where they explored federated learning and blockchain technology in a vehicular Ad Hoc Network (VANET) setting for smart transportation, focusing on security and privacy concerns. **Reputation-based distributed access control** architecture is recomended by Selcuk et al. in [141] that helps good peers create trust, identify and stop evil peers, and stop content propagation. Similarly, Tran et al. [142] proposed a dsitributed access control framework for P2P users.

Contrarily, distributed essential management-related methods are descriptively covered in [122] where authors focused on resource management, joint optimization, data management, compute offloading, and security mechanisms through the successful integration of blockchain and edge computing systems.

**Threshold signature** based distributed access control uses threshold cryptography for Collaboration-based access control where members approve a new member. At least a fraction of the group's members are applying. Access control lists (ACLs) cannot be used in a dynamic network because they are listed indefinitely. By discreetly sharing the signature key among group members and using a distributed protocol, members employ key sharing rather than the key itself. Threshold signatures can withstand system corruptions. Threshold signatures are employed in collaborative access control. In [143], Saxena et al. For example, a threshold signature based on the RSA (Rivest-Shamir-Adleman algorithm) [144], a threshold signature based on the DSA (Digital Signature Algorithm), a simple signature (PS), and Multi-signature Accountable Subgroup (ASM). Another DAC technique is **Trusted computing-based** technique which is only used in a real platform and programs in valid states, such as integrity and configuration, can access an item using trusted computing. The Trusted Reference Monitor (TRM) is an application-layer trusted computing architecture for access control for securing platform and software integrity and condition [145].

**List-based** DAC lists permissions which grant users and/or system processes access to and operations on items. Each access control list (ACL) item has a subject and an operation. Fenkam et al. [146] secure mobile P2P systems in collaborative settings. Peer-to-peer communication is possible by employing access control services on a mobile teamwork platform (MOTION). **Group signature-based** DAC allows a Peer groups which have no hierarchy and equal rights and obligations. A complex event-based IoT control framework [147] can be built utilizing tokens and smart contracts. Smart contracts [148] generate a Blockchain event whenever any function is called by a client. To let real clients use IoT devices, a smart contract-based blockchain solution built on Ethereum is used. In the system for controlling access and proving who you are, the Ethereum wallet address is used to check the client. The sender's access token and Ethereum address are only made public if the client is valid. IoT permissions can be granted using trustworthy oracles and the Blockchain. Service providers, Blockchain, and remote clients all communicate with one another via Oracle gateways [149]. In many applications uses customised DAC in combination with multiple access control mechanisms to increase security which is alos known as **Hybrid distributed access control**. For example, Lu et al. [150] incorporate hybrid DAC regulations into a single system based on the user's credentials, identity, and/or job. A non-centralized role-mapping technique is adopted [151]. Primary Backup (PB) fault-tolerant network job scheduling is utilized in multiprocessor system [152].

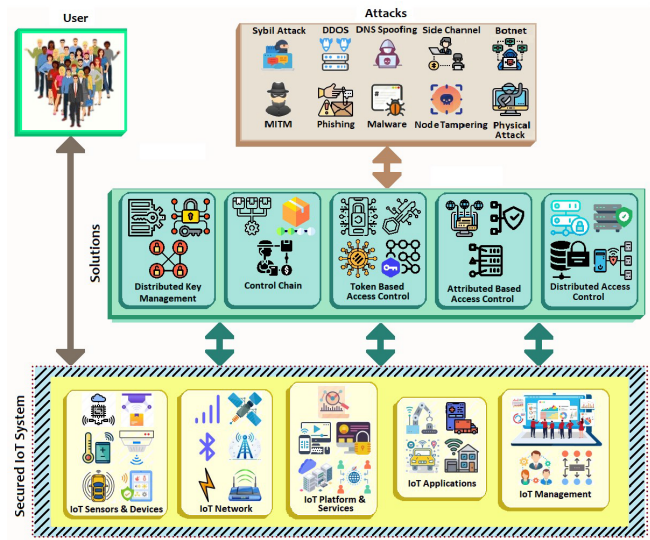

Fig. 7. IoT access control involves setting up policies and mechanisms that limit access to IoT devices and networks based on factors such as user identity, device identity, and security clearance.

## V. Attacks on Blockchain

Blockchain is renowned for its integrated security features and success statistics. However, there are some examples of attacks on existing Blockchain applications. This section illustrates some real-life attacks discussed in various research studies. Figure 8 describes layer-based Blockchain attacks, which encompass a broad category of attacks that aim to compromise various parts of the Blockchain protocol. The goal of these attacks is to weaken the Blockchain's defences by exploiting flaws in a particular layer. Attacks at the network layer, such as DDoS, can disrupt services and make consensus less reliable. Smart contract flaws can be exploited at the application layer. This includes attacks such as reentrancy attacks, in which the attacker manipulates the execution flow within a smart contract, potentially causing undesirable outcomes or financial losses.

### A. Consensus Related Attack

- **Double Spending Attacks:** The double-spending attack is one of the well-known attacks discussed in many articles like in [153]. Attacks involving double spending can also be dangerous in the context of the IoT. Blockchain technology is used in IoT systems to enable device interaction and value exchange. In public Blockchain, consensus process responsible to protect double-spending occurrences. In this instance, a decentralised network of users known as "miners" has performed the verification. Adding a transaction to the growing list of blocks means it has been checked and is legitimate. It becomes harder to reverse a transaction and spend the same money twice as more blocks are uploaded to the Blockchain[154].

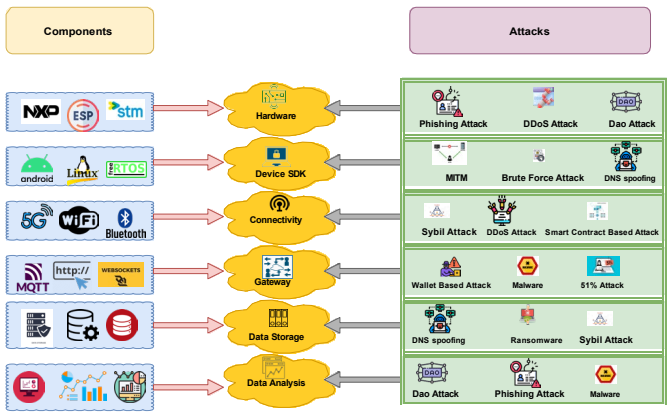

Fig. 8. Layer-based attacks in the Blockchain based IoT ecosystem

- **51% attack:** In the context of blockchain, a 51% attack occurs when one entity, or a collection of cooperating entities, controls more than 50% of the network's total processing power (hashrate) [155]. With this much power, they can influence the blockchain and maybe commit crimes like double spending or transaction censorship. A 51% attack can pose a serious risk to the integrity of blockchain-based IoT systems in the context of the IoT. Over 18 USD million in Bitcoin Gold (BTG) was double spent as a result of an assault in 2018 [156]. Similarly, over 100,000 USD worth of Vertcoin (VTC) was doubled in 2018 as a result of many attacks on the cryptocurrency [157]. A 2019 attack on Ethereum Classic (ETC) led to the double spending of more than 1 million USD of the cryptocurrency [158]. Three other attacks were also made against the cryptocurrency in 2020. Grin (GRIN) was the target of a 2020 attack, but the Blockchain was able to retake control [159]. In 2021, Bitcoin SV (BSV) was the target of three attacks that diminished its standing [160].

### B. Selfish Mining Attack

When one consensus participant or miner or a group of miners changes the Blockchain to help themselves, this is called "selfish mining." Miners "hide" their created blocks with this approach [161]. Bitcoin miners may be considered "Greedy" due to their pursuit of the block reward. While the selfish mining attack outlined here applies solely to evil miners, other miners are legitimate and reasonable [162]. Dishonest miners in selfish mining engage in data hiding and cause harm to other honest miners in two ways: (i) by receiving an unfair reward that is higher than the value of their offer of processing control spent [163], and (ii) by causing other miners to become confused and waste their resources on useless activities [164]. Since miners are what add new blocks to the Blockchain, this vulnerability affects everyone[165]. Valid blocks are withheld or invalid ones are added by some miners, making it harder to contribute real blocks [166].

.

### C. Smart contract-based attacks

The blockchain facilitates the operation of decentralized apps through smart contracts, which are self-executing lines of code. In the context of the IoT ecosystem, smart contract-based attacks on blockchain may target flaws in the smart contracts' source code or take advantage of their interactions with IoT devices [167]. As a result, once the smart contract has begun, it cannot be terminated. Furthermore, once a Blockchain-based transaction has been confirmed, it can no longer be altered. As a result, millions of dollars are at risk with no recourse if intelligent contracts are flawed. More than 3 million USD was stolen from The Tinyman exchange, which was built on the Algorand Blockchain in January 2022 [168]. In the Wormhole Cross Chain Bridge Attack of February 2022, almost 320 million USD were stolen from Ethereum and Solana [169]. In August 2022, about 8 million USD worth of Solana was lost when thousands of users experienced difficulties with account imports [170]. By August of 2021, hackers had stolen $613 million from a crypt by exploiting flaws in the Poly Network's smart contracts [171].

### D. Brute-Force Attack

A Brute-force attack is an improved version of the Finney attack in which an intelligent attacker takes command of n nodes in the network and uses them to execute a private mining scheme designed to carry out a Double-Spend assault [172]. In order to gain unauthorised access or affect the behaviour of the blockchain network, brute-force attacks against IoT-based blockchain systems often require making an attempt to guess or exhaustively testing several permutations of cryptographic keys or credentials. In a repeat of previous assaults, a malicious actor presents a double-spending transaction in a block while simultaneously expanding a private Blockchain [173]. The brute force attack and customer account breach at Dunkin' Donuts cost 650,000 USD in fines and damages and compelled the company to change all user passwords and increase application security [174]. There was no mention of monetary losses; however, nearly 20.6 million Alibaba accounts were hacked, and all users were instructed to change their passwords [175].

### E. Peer-to-peer network attacks

A Peer-to-peer (P2P) network where all participating peers run the protocol and have a copy of the transaction Ledger [176] allows for decentralized value exchanges via consensus. P2P network attacks against IoT devices inside a decentralised network can take advantage of communication and coordination flaws amongst IoT devices. The integrity, availability, and privacy of IoT devices may be jeopardised by these assaults, which have the potential to target different P2P network components. Following are a few instances of P2P network assaults in the IoT ecosystem and possible defences:

- **Sybil attack:** Sybil attacks occur when a single attacker assumes the identities of several targets. In the IoT ecosystem, Sybil attacks can lead to unauthorized access, data manipulation, or disruption of IoT device interactions. As such, it ranks high on the list of challenges one faces while trying to join a P2P system [177]. A Sybil attack and traffic confirmation attack were conducted against Tor in 2014 [178]. Other Sybil assaults on Tor users exist. This covers 2020 Bitcoin rewriting attacks. An attacker controlled a fourth of all Tor exit relays and used SSL stripping to downgrade secure connections and redirect payments to BTCMITM20's wallet [179]. Another example is KAX17's 2017-2021 attack. This group controlled 900 malicious servers, mostly midway points, to deanonymize Tor users [180].
- **Distributed denial-of-service:** A distributed denial-of-service (DDoS) assault is one in which numerous hacked computers work together to disrupt service for those attempting to access the attacked resource[124]. Targeting IoT devices in the P2P network, DDoS assaults overload them with a large volume of requests or malicious traffic, making them unreachable or interfering with their functioning. To disrupt legitimate users or systems, DDoS uses a barrage of incoming messages, connection requests, or corrupted packets to slow down or shut down the target system [181]. GitHub was hit by the largest DDoS attack ever in 2018 [182] [183]. Millions of programmers worldwide use the site to upload and distribute source code. In the year 2020, the New Zealand stock exchange was taken down by a massive distributed denial of service attack[184]. In 2019, China's Great Cannon DDoS operation clogged the internet by attacking a Hong Kong protest website. Hackers, hacktivists, and government-affiliated organizations use DDoS attacks in social movements. DDoS assaults can draw attention to a group or cause. Fancy Bear and Armada Collective have threatened DDoS attacks on numerous organizations in 2020. This combines DDoS with ransomware [185].

*F. Decentralised Autonomous Oragnisation (DAO) Attack*

DAO is an autonomous organization with no central leadership that codifies a company's internal procedures so that manual processes and human oversight of decentralized systems are unnecessary. This can be applied to IoT systems to create self-governing, transparent, decision making, data-sharing interoperability and autonomous networks of interconnected devices. Slock is an successful example of DAO application for Crowdfunding which was able to raise 12.7 million ether (about 150 million USD) [186]. However, a malicious actor discovers a flaw in the program. Thus, the attacker begins the attack by depositing a little amount and then using a recursive withdraw method to extract 70 million USD [187]. The Ethereum Foundation issued a warning to a potential attacker, threatening to freeze their account until they stopped participating in the contract or agreed to a hard fork. A while later, he put a halt to the assault. This attack on a DAO has generated many future worries about the autonomy of smart contracts [188].

*G. Wallet-based attacks*

The wallet automates payments in BC which is executed through Network nodes payment processing. A wallet-based attack in the context of IoT systems typically refers to an attack targeting the digital wallets or cryptocurrency wallets associated with IoT devices. A Blockchain wallet may be targeted if a node executes a malicious behavior [189]. In an instance of responsibility, an attacker took 77 million USD from a parity client's wallet, which required multi-sign verification [190]. To reduce transaction fees, the multi-sign party wallet employs a library contract. An attacker exploited security flaws in this functionality [191]. When the attacker listed his account as the library's owner, he gained equal ownership of all wallets [192]. Next, the attacker alters the library code that froze the currency.

Blockchain is widely popular for cryptocurrencies where transactions are not continuous. For instance, when the sender sends and the recipient receives, the transaction is done. Contrarily, transactions in the IoT ecosystem are typically continuous even when their outcomes are used as a separate parameter to control other devices [10]. For instance, when a fire alarm sounds, an automatic tap will open as soon as the smoke detector responds. As a result, it differs greatly from financial transactions, which call for quick choices. While blockchain technology has many advantages for IoT access control, it is still in its infancy and cannot resolve all complex problems. Scalability, for instance, will be difficult as IoT devices and transactions grow in quantity [17], [113]. This might strain the blockchain network, resulting in slower transaction processing times and higher costs. In addition, massive IoT devices are being connected throughout the ecosystem to handle enormous volumes of transactions. As a result, their consensus approval, and fitting transactions into ledgers require a lot of bandwidth, which also increases latency [193]. Implementing a time-sensitive, blockchain-based IoT ecosystem could be exceedingly difficult due to this latency. This latency can be a drawback in the context of IoT access control, where real-time interactions are essential.

## VI. FUTURE RESEARCH DIRECTION

In fact, both the academic community and the business world pay close attention to blockchain. It is still not developed enough to be adopted in all application domains, despite being a ten-year-old technology. Large IT corporations, including IBM, Intel, and others, have already launched their blockchain-based service platforms for a variety of small and medium-sized businesses, but independent service providers are still far behind. As a result, there are several chances for this technology to be developed through study and commercial use. This section focuses on a few of the potential avenues 4for future research that we have identified from various recent research contributions.

- **Performance and scalability:** Scalability of blockchains is still an important subject for research. In order to increase the scalability and performance of blockchain-based IoT systems, researchers should look into methods like sharding, off-chain transactions, and layer-two solutions [17]. It's also critical to research efficient consensus techniques for low-resource IoT devices.
- **Privacy and confidentiality:** It's crucial to improve privacy and secrecy in blockchain-based IoT systems. IoT scenarios can benefit from the investigation of techniques like homomorphic encryption, zero-knowledge proofs, and secure multiparty computation. To safeguard the identity and operations of IoT devices, research can be done on anonymization and unlinkability technologies.
- **Interoperability for Automation:** To enable seamless communication and collaboration targeting smart automation between various IoT devices and blockchain networks, interoperability frameworks and standards for blockchain-based IoT systems must be researched . The full potential of decentralised IoT ecosystems can be unlocked by creating protocols and designs that enable interoperability across multiple blockchain platforms and IoT protocols [114].
- **Governance and consensus:** For blockchain-based IoT systems, investigating efficient governance models and incentive structures can guarantee equitable and effective decision-making, resource distribution, and participation. Creating incentives for device owners, users, and other stakeholders to take part in the blockchain network and aid in its growth is essential. Study governance models for private blockchains, including consensus algorithms and decision-making processes [193]. Research can focus on designing consensus mechanisms that are efficient, secure, and suitable for the specific requirements of private blockchain networks.
- **Security and robustness:** It is necessary to conduct research on cutting-edge cryptographic methods, secure key management, and secure hardware integration to improve trust and security in blockchain-based IoT systems. Important research areas include exploring secure firmware upgrades and safe bootstrapping procedures for IoT devices, as well as investigating techniques to detect and mitigate assaults like Sybil and 51% attacks [194].
- **Integration with Edge Computing:** Investigating the integration of blockchain with edge computing can bring computation and storage capabilities closer to IoT devices, reducing latency and improving responsiveness. Blockchain-based IoT systems can be effectively implemented through research on the distributed execution of smart contracts at the edge, secure data sharing between edge nodes, and optimised resource management.
- **Real-World Applications:** Research on actual use cases and applications of blockchain-based IoT systems might shed light on their advantages and practical difficulties . Finding specific requirements, creating best practises, and evaluating the impact of blockchain technology in various contexts can all be aided by investigating areas like supply chain management, energy grids, healthcare, and smart cities [195].

## VII. Conclusion

This article examines the IoT ecosystem's general security concerns and potential remedies. To make things more apparent, we have concentrated on three crucial factors: the most common threats, security challenges and their current remedies, particularly in access control, and finally, blockchain-based solutions. We have summarised recent related surveys and the gap that we have filled in this article. For example, most of the recent contributions identify access control as one of the key challenges in the IoT ecosystem. In addition, a huge number of them suggested blockchain as a potential solution. Considering this fact, we reviewed the majority of the access control challenges and considered the articles that contributed to blockchain as potential solutions. We focus on every technical aspect that can impact an IoT system's security. As blockchain is widely considered, we further tried to summarise recent track records of attacks on blockchain. We believe proper technical analysis of these attacks can aid academics in identifying research gaps in the IoT ecosystem. The research gaps in this area have yet to be filled, which we listed in future directions for further research. The researchers who are working in this field will be guided by the direction of future research. This essay wraps up recent ideas on guaranteeing access control in the IoT ecosystem.

## References

[1] J. Holler, V. Tsiatsis, C. Mulligan, S. Karnouskos, S. Avesand, and D. Boyle, *Internet of Things*. Academic Press, 2014.
[2] M. M. Hasan, M. A. Rahman, A. Sedigh, A. U. Khasanah, A. T. Asyhari, H. Tao, and S. A. Bakar, "Search and rescue operation in flooded areas: A survey on emerging sensor networking-enabled iot-oriented technologies and applications," *Cognitive Systems Research*, vol. 67, pp. 104–123, 2021.
[3] S. Al-Sarawi, M. Anbar, R. Abdullah, and A. B. Al Hawari, "Internet of things market analysis forecasts, 2020–2030," in *2020 Fourth World Conference on smart trends in systems, security and sustainability (WorldS4)*, pp. 449–453, IEEE, 2020.
[4] S. Biswas, Z. Yao, L. Yan, A. Alqhatani, A. K. Bairagi, F. Asiri, and M. Masud, "Interoperability benefits and challenges in smart city services: Blockchain as a solution," *Electronics*, vol. 12, no. 4, 2023.
[5] H. R. Lim, K. S. Khoo, W. Y. Chia, K. W. Chew, S.-H. Ho, and P. L. Show, "Smart microalgae farming with internet-of-things for sustainable agriculture," *Biotechnology Advances*, p. 107931, 2022.
[6] S. P. Ramu, P. Boopalan, Q.-V. Pham, P. K. R. Maddikunta, T. Huynh-The, M. Alazab, T. T. Nguyen, and T. R. Gadekallu, "Federated learning enabled digital twins for smart cities: Concepts, recent advances, and future directions," *Sustainable Cities and Society*, vol. 79, p. 103663, 2022.
[7] C. Perera, C. H. Liu, and S. Jayawardena, "The emerging internet of things marketplace from an industrial perspective: A survey," *IEEE transactions on emerging topics in computing*, vol. 3, no. 4, pp. 585–598, 2015.
[8] M. H. Kashani, M. Madanipour, M. Nikravan, P. Asghari, and E. Mahdipour, "A systematic review of iot in healthcare: Applications, techniques, and trends," *Journal of Network and Computer Applications*, vol. 192, p. 103164, 2021.
[9] B. Nour, K. Sharif, F. Li, S. Biswas, H. Moungla, M. Guizani, and Y. Wang, "A survey of internet of things communication using icn: A use case perspective," *Computer Communications*, vol. 142-143, pp. 95–123, 2019.

[10] S. Biswas, K. Sharif, F. Li, and S. Mohanty, "Blockchain for e-healthcare systems: Easier said than done," *Computer*, vol. 53, no. 7, pp. 57–67, 2020.

[11] W. Ejaz and A. Anpalagan, *Blockchain Technology for Security and Privacy in Internet of Things*, pp. 47–55. Cham: Springer International Publishing, 2019.

[12] T. Dalgleish, J. M. G. Williams, A.-M. J. Golden, N. Perkins, L. F. Barrett, P. J. Barnard, C. A. Yeung, V. Murphy, R. Elward, K. Tchanturia, *et al.*, "The blockchain-enabled intelligent iot economy," *J Exp Psychol Gen*, vol. 136, pp. 23–42, 2018.

[13] A. Javaid, Q. Niyaz, W. Sun, and M. Alam, "A deep learning approach for network intrusion detection system," *Eai Endorsed Transactions on Security and Safety*, vol. 3, no. 9, p. e2, 2016.

[14] M. A. Ferrag, M. Derdour, M. Mukherjee, A. Derhab, L. Maglaras, and H. Janicke, "Blockchain technologies for the internet of things: Research issues and challenges," *IEEE Internet of Things Journal*, vol. 6, no. 2, pp. 2188–2204, 2018.

[15] L. Cui, S. Yang, Z. Chen, Y. Pan, M. Xu, and K. Xu, "An efficient and compacted dag-based blockchain protocol for industrial internet of things," *IEEE Transactions on Industrial Informatics*, vol. 16, no. 6, pp. 4134–4145, 2019.

[16] K.-K. R. Choo, Z. Yan, and W. Meng, "Blockchain in industrial iot applications: security and privacy advances, challenges, and opportunities," *IEEE Transactions on Industrial Informatics*, vol. 16, no. 6, pp. 4119–4121, 2020.

[17] S. Biswas, K. Sharif, F. Li, B. Nour, and Y. Wang, "A scalable blockchain framework for secure transactions in iot," *IEEE Internet of Things Journal*, vol. 6, no. 3, pp. 4650–4659, 2019.

[18] S. Biswas, K. Sharif, F. Li, I. Alam, and S. P. Mohanty, "Daac: Digital asset access control in a unified blockchain based e-health system," *IEEE Transactions on Big Data*, vol. 8, no. 5, pp. 1273–1287, 2022.

[19] M. Sookhak, M. R. Jabbarpour, N. S. Safa, and F. R. Yu, "Blockchain and smart contract for access control in healthcare: a survey, issues and challenges, and open issues," *Journal of Network and Computer Applications*, vol. 178, p. 102950, 2021.

[20] F. Ghaffari, E. Bertin, J. Hatin, and N. Crespi, "Authentication and access control based on distributed ledger technology: a survey," in *2020 2nd Conference on Blockchain Research & Applications for Innovative Networks and Services (BRAINS)*, pp. 79–86, IEEE, 2020.

[21] P. Bagga, A. K. Das, V. Chamola, and M. Guizani, "Blockchain-envisioned access control for internet of things applications: a comprehensive survey and future directions," *Telecommunication Systems*, pp. 1–49, 2022.

[22] S. Pal, A. Dorri, and R. Jurdak, "Blockchain for iot access control: Recent trends and future research directions," *Journal of Network and Computer Applications*, p. 103371, 2022.

[23] M. Shakarami, J. Benson, and R. Sandhu, "Blockchain-based administration of access in smart home iot," in *Proceedings of the 2022 ACM Workshop on Secure and Trustworthy Cyber-Physical Systems*, pp. 57–66, 2022.

[24] R. Sekaran, R. Patan, A. Raveendran, F. Al-Turjman, M. Ramachandran, and L. Mostarda, "Survival study on blockchain based 6g-enabled mobile edge computation for iot automation," *IEEE access*, vol. 8, pp. 143453–143463, 2020.

[25] M. Abubakar, Z. Jaroucheh, A. Al Dubai, and B. Buchanan, "A decentralised authentication and access control mechanism for medical wearable sensors data," in *2021 IEEE International Conference on Omni-Layer Intelligent Systems (COINS)*, pp. 1–7, IEEE, 2021.

[26] L. Bai, M. Hu, M. Liu, and J. Wang, "Bpiiot: A light-weighted blockchain-based platform for industrial iot," *IEEE Access*, vol. 7, pp. 58381–58393, 2019.

[27] F. Wang, Z. Hu, H. Wang, X. Chen, and W. Feng, "Cross-domain dynamic access control based on "blockchain+ artificial intelligence"," *Neural Computing and Applications*, pp. 1–11, 2023.

[28] A. Yazdinejad, A. Dehghantanha, R. M. Parizi, G. Srivastava, and H. Karimipour, "Secure intelligent fuzzy blockchain framework: Effective threat detection in iot networks," *Computers in Industry*, vol. 144, p. 103801, 2023.

[29] K. H. Y. Chung, D. Li, and P. Adriaens, "Technology-enabled financing of sustainable infrastructure: A case for blockchains and decentralized oracle networks," *Technological Forecasting and Social Change*, vol. 187, p. 122258, 2023.

[30] M. S. Mahmood and N. B. Al Dabagh, "Blockchain technology and internet of things: review, challenge and security concern," *International Journal of Electrical and Computer Engineering*, vol. 13, no. 1, p. 718, 2023.

[31] Originstamp, "Benefits and challenges of blockchain in iot," 2021.

[32] B. K. Mohanta, D. Jena, U. Satapathy, and S. Patnaik, "Survey on iot security: Challenges and solution using machine learning, artificial intelligence and blockchain technology," *Internet of Things*, vol. 11, p. 100227, 2020.

[33] X. Wang, X. Zha, W. Ni, R. P. Liu, Y. J. Guo, X. Niu, and K. Zheng, "Survey on blockchain for internet of things," *Computer Communications*, vol. 136, pp. 10–29, 2019.

[34] R. A. Memon, J. P. Li, J. Ahmed, M. I. Nazeer, M. Ismail, and K. Ali, "Cloud-based vs. blockchain-based iot: a comparative survey and way forward," *Frontiers of Information Technology & Electronic Engineering*, vol. 21, no. 4, pp. 563–586, 2020.

[35] W. Gao, W. G. Hatcher, and W. Yu, "A survey of blockchain: Techniques, applications, and challenges," in *2018 27th international conference on computer communication and networks (ICCCN)*, pp. 1–11, IEEE, 2018.

[36] H.-N. Dai, Z. Zheng, and Y. Zhang, "Blockchain for internet of things: A survey," *IEEE Internet of Things Journal*, vol. 6, no. 5, pp. 8076–8094, 2019.

[37] M. A. Uddin, A. Stranieri, I. Gondal, and V. Balasubramanian, "A survey on the adoption of blockchain in iot: Challenges and solutions," *Blockchain: Research and Applications*, p. 100006, 2021.

[38] F. Alkurdi, I. Elgendi, K. S. Munasinghe, D. Sharma, and A. Jamalipour, "Blockchain in iot security: a survey," in *2018 28th International Telecommunication Networks and Applications Conference (ITNAC)*, pp. 1–4, IEEE, 2018.

[39] A. Panarello, N. Tapas, G. Merlino, F. Longo, and A. Puliafito, "Blockchain and iot integration: A systematic survey," *Sensors*, vol. 18, no. 8, p. 2575, 2018.

[40] E. F. Jesus, V. R. Chicarino, C. V. De Albuquerque, and A. A. d. A. Rocha, "A survey of how to use blockchain to secure internet of things and the stalker attack," *Security and Communication Networks*, vol. 2018, 2018.

[41] M. Salimitari and M. Chatterjee, "A survey on consensus protocols in blockchain for iot networks," *arXiv preprint arXiv:1809.05613*, 2018.

[42] Q. Feng, D. He, S. Zeadally, M. K. Khan, and N. Kumar, "A survey on privacy protection in blockchain system," *Journal of Network and Computer Applications*, vol. 126, pp. 45–58, 2019.

[43] M. Wu, K. Wang, X. Cai, S. Guo, M. Guo, and C. Rong, "A comprehensive survey of blockchain: From theory to iot applications and beyond," *IEEE Internet of Things Journal*, vol. 6, no. 5, pp. 8114–8154, 2019.

[44] R. Thakore, R. Vaghashiya, C. Patel, and N. Doshi, "Blockchain-based iot: A survey," *Procedia Computer Science*, vol. 155, pp. 704–709, 2019.

[45] J. Zhang, S. Zhong, T. Wang, H.-C. Chao, and J. Wang, "Blockchain-based systems and applications: a survey," *Journal of Internet Technology*, vol. 21, no. 1, pp. 1–14, 2020.

[46] A. Al Sadawi, M. S. Hassan, and M. Ndiaye, "A survey on the integration of blockchain with iot to enhance performance and eliminate challenges," *IEEE Access*, vol. 9, pp. 54478–54497, 2021.

[47] L. Da Xu, Y. Lu, and L. Li, "Embedding blockchain technology into iot for security: a survey," *IEEE Internet of Things Journal*, 2021.

[48] S. Namane and I. Ben Dhaou, "Blockchain-based access control techniques for iot applications," *Electronics*, vol. 11, no. 14, p. 2225, 2022.

[49] P. Sharma, R. Jindal, and M. D. Borah, "Blockchain-based cloud storage system with cp-abe-based access control and revocation process," *The Journal of Supercomputing*, pp. 1–29, 2022.

[50] M. S. Rahman, M. A. Islam, M. A. Uddin, and G. Stea, "A survey of blockchain-based iot ehealthcare: Applications, research issues, and challenges," *Internet of Things*, vol. 19, p. 100551, 2022.

[51] J. Han, Y. Zhang, J. Liu, Z. Li, M. Xian, H. Wang, F. Mao, and Y. Chen, "A blockchain-based and sgx-enabled access control framework for iot," *Electronics*, vol. 11, no. 17, p. 2710, 2022.

[52] M. M. Ahemd, M. A. Shah, and A. Wahid, "Iot security: A layered approach for attacks & defenses," in *2017 international conference on Communication Technologies (ComTech)*, pp. 104–110, IEEE, 2017.

[53] L. C. Mutalemwa and S. Shin, "Comprehensive performance analysis of privacy protection protocols utilizing fake packet injection techniques," *IEEE Access*, vol. 8, pp. 76935–76950, 2020.

[54] H. Pirayesh, P. K. Sangdeh, and H. Zeng, "Securing zigbee communications against constant jamming attack using neural network," *IEEE Internet of Things Journal*, vol. 8, no. 6, pp. 4957–4968, 2020.

[55] A. Bose, G. S. Aujla, M. Singh, N. Kumar, and H. Cao, "Blockchain as a service for software defined networks: A denial of service attack perspective," in *2019 IEEE Intl Conf on Dependable, Autonomic and Secure Computing, Intl Conf on Pervasive Intelli-*

*gence and Computing, Intl Conf on Cloud and Big Data Computing, Intl Conf on Cyber Science and Technology Congress (DASC/PiCom/CBDCom/CyberSciTech)*, pp. 901–906, IEEE, 2019.

[56] I. Andrea, C. Chrysostomou, and G. Hadjichristofi, "Internet of things: Security vulnerabilities and challenges," in *2015 IEEE Symposium on Computers and Communication (ISCC)*, pp. 180–187, 2015.

[57] E. Stancu, M. Berceanu, S. Halunga, and O. Fratu, "Rfid sensitivity in narrowband jamming environment," in *Advanced Topics in Optoelectronics, Microelectronics, and Nanotechnologies XI*, vol. 12493, pp. 699–703, SPIE, 2023.

[58] R. Sahay, G. Geethakumari, and B. Mitra, "A novel blockchain based framework to secure iot-llns against routing attacks," *Computing*, vol. 102, no. 11, pp. 2445–2470, 2020.

[59] A. McCarthy, E. Ghadafi, P. Andriotis, and P. Legg, "Defending against adversarial machine learning attacks using hierarchical learning: A case study on network traffic attack classification," *Journal of Information Security and Applications*, vol. 72, p. 103398, 2023.

[60] S. Luo, Y. Lai, and J. Liu, "Selective forwarding attack detection and network recovery mechanism based on cloud-edge cooperation in software-defined wireless sensor network," *Computers & Security*, vol. 126, p. 103083, 2023.

[61] P. Varga, S. Plosz, G. Soos, and C. Hegedus, "Security threats and issues in automation iot," in *2017 IEEE 13th International Workshop on Factory Communication Systems (WFCS)*, pp. 1–6, IEEE, 2017.

[62] A. La Salle, A. Kumar, P. Jevtic´, and D. Boscovic, "Joint modeling of hyperledger fabric and sybil attack: Petri net approach," *Simulation Modelling Practice and Theory*, vol. 122, p. 102674, 2023.

[63] S. Biswas, K. Sharif, F. Li, and Y. Liu, "3p framework: Customizable permission architecture for mobile applications," in *Wireless Algorithms, Systems, and Applications* (L. Ma, A. Khreishah, Y. Zhang, and M. Yan, eds.), (Cham), pp. 445–456, Springer International Publishing, 2017.

[64] S. Biswas, W. Haipeng, and J. Rashid, "Android Permissions Management at App Installing," *International Journal of Security and Its Applications*, pp. 223–232, mar 10 2016.

[65] A. B. de Neira, B. Kantarci, and M. Nogueira, "Distributed denial of service attack prediction: Challenges, open issues and opportunities," *Computer Networks*, p. 109553, 2023.

[66] R. Singh, S. Tanwar, and T. P. Sharma, "Utilization of blockchain for mitigating the distributed denial of service attacks," *Security and Privacy*, vol. 3, no. 3, p. e96, 2020.

[67] A. Buriro, A. B. Buriro, T. Ahmad, S. Buriro, and S. Ullah, "Malwd&c: A quick and accurate machine learning-based approach for malware detection and categorization," *Applied Sciences*, vol. 13, no. 4, p. 2508, 2023.

[68] X. Ling, L. Wu, J. Zhang, Z. Qu, W. Deng, X. Chen, Y. Qian, C. Wu, S. Ji, T. Luo, *et al.*, "Adversarial attacks against windows pe malware detection: A survey of the state-of-the-art," *Computers & Security*, p. 103134, 2023.

[69] G. A. Abdalrahman and H. Varol, "Defending against cyber-attacks on the internet of things," in *2019 7th International Symposium on Digital Forensics and Security (ISDFS)*, pp. 1–6, IEEE, 2019.

[70] A. C. B. Monteiro, R. P. Franc¸a, R. Arthur, and Y. Iano, "The fundamentals and potential for cyber security of machine learning in the modern world," in *Advanced Smart Computing Technologies in Cybersecurity and Forensics*, pp. 119–137, CRC Press, 2021.

[71] A. Dua, V. Tyagi, N. Patel, and B. Mehtre, "Iisr: A secure router for iot networks," in *2019 4th international conference on information systems and computer networks (ISCON)*, pp. 636–643, IEEE, 2019.

[72] I. Alam, K. Sharif, F. Li, Z. Latif, M. M. Karim, S. Biswas, B. Nour, and Y. Wang, "A survey of network virtualization techniques for internet of things using sdn and nfv," vol. 53, apr 2020.

[73] J. Sengupta, S. Ruj, and S. Dasbit, "A comprehensive survey on attacks, security issues and blockchain solutions for iot and iiot," *Journal of Network and Computer Applications*, 11 2019.

[74] P. Shaverdian, "Start with trust: utilizing blockchain to resolve the third-party data breach problem," *UCLA L. Rev.*, vol. 66, p. 1242, 2019.

[75] W. Feng, Y. Li, X. Yang, Z. Yan, and L. Chen, "Blockchain-based data transmission control for tactical data link," *Digital Communications and Networks*, vol. 7, no. 3, pp. 285–294, 2021.

[76] C. DeCusatis, M. Zimmermann, and A. Sager, "Identity-based network security for commercial blockchain services," in *2018 IEEE 8th Annual Computing and Communication Workshop and Conference (CCWC)*, pp. 474–477, IEEE, 2018.

[77] B. Mohanta, U. Satapathy, S. Panda, and D. Jena, "A novel approach to solve security and privacy issues for iot applications using blockchain," pp. 394–399, 12 2019.

[78] M. A. Ferrag, L. Shu, X. Yang, A. Derhab, and L. Maglaras, "Security and privacy for green iot-based agriculture: Review, blockchain solutions, and challenges," *IEEE access*, vol. 8, pp. 32031–32053, 2020.

[79] N. Waheed, X. He, M. Ikram, M. Usman, S. S. Hashmi, and M. Usman, "Security and privacy in iot using machine learning and blockchain: Threats and countermeasures," *ACM Computing Surveys (CSUR)*, vol. 53, no. 6, pp. 1–37, 2020.

[80] Y. Yu, Y. Li, J. Tian, and J. Liu, "Blockchain-based solutions to security and privacy issues in the internet of things," *IEEE Wireless Communications*, vol. 25, no. 6, pp. 12–18, 2018.

[81] N. M. Kumar and P. K. Mallick, "Blockchain technology for security issues and challenges in iot," *Procedia Computer Science*, vol. 132, pp. 1815–1823, 2018.

[82] S. N. Mohanty, K. Ramya, S. S. Rani, D. Gupta, K. Shankar, S. Lakshmanaprabu, and A. Khanna, "An efficient lightweight integrated blockchain (elib) model for iot security and privacy," *Future Generation Computer Systems*, vol. 102, pp. 1027–1037, 2020.

[83] S.-Y. Kuo, F.-H. Tseng, and Y.-H. Chou, "Metaverse intrusion detection of wormhole attacks based on a novel statistical mechanism," *Future Generation Computer Systems*, vol. 143, pp. 179–190, 2023.

[84] S. M. Nair, V. Ramesh, and A. K. Tyagi, "Issues and challenges (privacy, security, and trust) in blockchain-based applications," in *Research Anthology on Convergence of Blockchain, Internet of Things, and Security*, pp. 1101–1114, IGI Global, 2023.

[85] K. Hameed, M. Barika, S. Garg, M. B. Amin, and B. Kang, "A taxonomy study on securing blockchain-based industrial applications: An overview, application perspectives, requirements, attacks, countermeasures, and open issues," *Journal of Industrial Information Integration*, vol. 26, p. 100312, 2022.

[86] B. S. Egala, A. K. Pradhan, V. Badarla, and S. P. Mohanty, "Fortified-chain: a blockchain-based framework for security and privacy-assured internet of medical things with effective access control," *IEEE Internet of Things Journal*, vol. 8, no. 14, pp. 11717–11731, 2021.

[87] D. Gupta, S. Bhatt, M. Gupta, O. Kayode, and A. S. Tosun, "Access control model for google cloud iot," in *2020 IEEE 6th Intl Conference on Big Data Security on Cloud (BigDataSecurity), IEEE Intl Conference on High Performance and Smart Computing,(HPSC) and IEEE Intl Conference on Intelligent Data and Security (IDS)*, pp. 198–208, IEEE, 2020.

[88] M. El-Hajj, A. Fadlallah, M. Chamoun, and A. Serhrouchni, "A survey of internet of things (iot) authentication schemes," *Sensors*, vol. 19, no. 5, p. 1141, 2019.

[89] N. Tapas, G. Merlino, and F. Longo, "Blockchain-based iot-cloud authorization and delegation," in *2018 IEEE International Conference on Smart Computing (SMARTCOMP)*, pp. 411–416, IEEE, 2018.

[90] N. Bhalaji, "Reliable data transmission with heightened confidentiality and integrity in iot empowered mobile networks," *Journal of ISMAC*, vol. 2, no. 02, pp. 106–117, 2020.

[91] N. F. Syed, Z. Baig, A. Ibrahim, and C. Valli, "Denial of service attack detection through machine learning for the iot," *Journal of Information and Telecommunication*, vol. 4, no. 4, pp. 482–503, 2020.

[92] X. Yu, J. Qiu, X. Yang, Y. Cong, and L. Du, "An graph-based adaptive method for fast detection of transformed data leakage in iot via wsn," *IEEE Access*, vol. 7, pp. 137111–137121, 2019.

[93] S. Andy, B. Rahardjo, and B. Hanindhito, "Attack scenarios and security analysis of mqtt communication protocol in iot system," in *2017 4th international conference on electrical engineering, computer science and informatics (EECSI)*, pp. 1–6, IEEE, 2017.

[94] S. J. Lee, P. D. Yoo, A. T. Asyhari, Y. Jhi, L. Chermak, C. Y. Yeun, and K. Taha, "Impact: Impersonation attack detection via edge computing using deep autoencoder and feature abstraction," *IEEE Access*, vol. 8, pp. 65520–65529, 2020.

[95] K. Hofer-Schmitz and B. Stojanovic´, "Towards formal verification of iot protocols: A review," *Computer Networks*, vol. 174, p. 107233, 2020.

[96] H. Si, C. Sun, Y. Li, H. Qiao, and L. Shi, "Iot information sharing security mechanism based on blockchain technology," *Future generation computer systems*, vol. 101, pp. 1028–1040, 2019.

[97] M. Wazid, P. Reshma Dsouza, A. K. Das, V. Bhat K, N. Kumar, and J. J. Rodrigues, "Rad-ei: a routing attack detection scheme for edge-based internet of things environment," *International Journal of Communication Systems*, vol. 32, no. 15, p. e4024, 2019.

[98] Y. Xu, J. Ren, G. Wang, C. Zhang, J. Yang, and Y. Zhang, "A blockchain-based nonrepudiation network computing service scheme for industrial iot," *IEEE Transactions on Industrial Informatics*, vol. 15, no. 6, pp. 3632–3641, 2019.

[99] R. Vishwakarma and A. K. Jain, "A survey of ddos attacking techniques and defence mechanisms in the iot network," *Telecommunication systems*, vol. 73, no. 1, pp. 3–25, 2020.

[100] D. Pavithran, K. Shaalan, J. N. Al-Karaki, and A. Gawanmeh, "Towards building a blockchain framework for iot," *Cluster Computing*, vol. 23, no. 3, pp. 2089–2103, 2020.

[101] J. C. Song, M. A. Demir, J. J. Prevost, and P. Rad, "Blockchain design for trusted decentralized iot networks," in *2018 13th Annual Conference on System of Systems Engineering (SoSE)*, pp. 169–174, IEEE, 2018.

[102] B. Alotaibi, "Utilizing blockchain to overcome cyber security concerns in the internet of things: A review," *IEEE Sensors Journal*, vol. 19, no. 23, pp. 10953–10971, 2019.

[103] M. Hamad, A. Finkenzeller, H. Liu, J. Lauinger, V. Prevelakis, and S. Steinhorst, "Seemqtt: Secure end-to-end mqtt-based communication for mobile iot systems using secret sharing and trust delegation," *IEEE Internet of Things Journal*, 2022.

[104] S. Parween and S. Z. Hussain, "A comparative analysis of coap based congestion control in iot," in *2021 4th International Conference on Recent Trends in Computer Science and Technology (ICRTCST)*, pp. 321–324, IEEE, 2022.

[105] M. Lima, R. Lima, F. Lins, and M. Bonfim, "Beholder–a cep-based intrusion detection and prevention systems for iot environments," *Computers & Security*, vol. 120, p. 102824, 2022.

[106] V. K. Quy, N. V. Hau, D. V. Anh, and L. A. Ngoc, "Smart healthcare iot applications based on fog computing: architecture, applications and challenges," *Complex & Intelligent Systems*, vol. 8, no. 5, pp. 3805–3815, 2022.

[107] I. Martins, J. S. Resende, P. R. Sousa, S. Silva, L. Antunes, and J. Gama, "Host-based ids: A review and open issues of an anomaly detection system in iot," *Future Generation Computer Systems*, 2022.

[108] C. Wan, A. Mehmood, M. Carsten, G. Epiphaniou, and J. Lloret, "A blockchain based forensic system for iot sensors using mqtt protocol," in *2022 9th International Conference on Internet of Things: Systems, Management and Security (IOTSMS)*, pp. 1–8, IEEE, 2022.

[109] W. Xiang and Z. Yuanyuan, "Scalable access control scheme of internet of things based on blockchain," *Procedia Computer Science*, vol. 198, pp. 448–453, 2022.

[110] T. A. Ahanger, A. Aljumah, and M. Atiquzzaman, "State-of-the-art survey of artificial intelligent techniques for iot security," *Computer Networks*, p. 108771, 2022.

[111] A. Pandey, R. Vamsi, and S. Kumar, "Handling device heterogeneity and orientation using multistage regression for gmm based localization in iot networks," *IEEE Access*, vol. 7, pp. 144354–144365, 2019.

[112] A. L. M. Neto, L. Barbosa, Í. F. S. Cunha, *et al.*, "Authentication of things: Authentication and access control for the entire iot device lifecycle," in *Anais Estendidos do XXII Simpósio Brasileiro em Segurança da Informação e de Sistemas Computacionais*, pp. 49–56, SBC, 2022.

[113] H. Cui, X. Yi, and S. Nepal, "Achieving scalable access control over encrypted data for edge computing networks," *IEEE Access*, vol. 6, pp. 30049–30059, 2018.

[114] S. Biswas, K. Sharif, F. Li, Z. Latif, S. S. Kanhere, and S. P. Mohanty, "Interoperability and synchronization management of blockchain-based decentralized e-health systems," *IEEE Transactions on Engineering Management*, vol. 67, no. 4, pp. 1363–1376, 2020.

[115] M. Gupta and R. Sandhu, "Towards activity-centric access control for smart collaborative ecosystems," in *Proceedings of the 26th ACM Symposium on Access Control Models and Technologies*, pp. 155–164, 2021.

[116] X. Yao, F. Farha, R. Li, I. Psychoula, L. Chen, and H. Ning, "Security and privacy issues of physical objects in the iot: Challenges and opportunities," *Digital Communications and Networks*, vol. 7, no. 3, pp. 373–384, 2021.

[117] A. Iftekhar, X. Cui, Q. Tao, and C. Zheng, "Hyperledger fabric access control system for internet of things layer in blockchain-based applications," *Entropy*, vol. 23, no. 8, p. 1054, 2021.

[118] P. Patil, M. Sangeetha, and V. Bhaskar, "Blockchain for iot access control, security and privacy: a review," *Wireless Personal Communications*, vol. 117, no. 3, pp. 1815–1834, 2021.

[119] Y. Feng, W. Zhang, X. Luo, and B. Zhang, "A consortium blockchain-based access control framework with dynamic orderer node selection for 5g-enabled industrial iot," *IEEE Transactions on Industrial Informatics*, vol. 18, no. 4, pp. 2840–2848, 2021.

[120] L. Tan, N. Shi, K. Yu, M. Aloqaily, and Y. Jararweh, "A blockchain-empowered access control framework for smart devices in green internet of things," *ACM Transactions on Internet Technology (TOIT)*, vol. 21, no. 3, pp. 1–20, 2021.

[121] N. Deepa, Q.-V. Pham, D. C. Nguyen, S. Bhattacharya, B. Prabadevi, T. R. Gadekallu, P. K. R. Maddikunta, F. Fang, and P. N. Pathirana, "A survey on blockchain for big data: approaches, opportunities, and future directions," *Future Generation Computer Systems*, 2022.

[122] H. Xue, D. Chen, N. Zhang, H.-N. Dai, and K. Yu, "Integration of blockchain and edge computing in internet of things: A survey," *arXiv preprint arXiv:2205.13160*, 2022.

[123] A. R. Javed, M. A. Hassan, F. Shahzad, W. Ahmed, S. Singh, T. Baker, and T. R. Gadekallu, "Integration of blockchain technology and federated learning in vehicular (iot) networks: A comprehensive survey," *Sensors*, vol. 22, no. 12, p. 4394, 2022.

[124] Z. Shah, I. Ullah, H. Li, A. Levula, and K. Khurshid, "Blockchain based solutions to mitigate distributed denial of service (ddos) attacks in the internet of things (iot): A survey," *Sensors*, vol. 22, no. 3, p. 1094, 2022.

[125] D. Pennino, M. Pizzonia, A. Vitaletti, and M. Zecchini, "Blockchain as iot economy enabler: A review of architectural aspects," *Journal of Sensor and Actuator Networks*, vol. 11, no. 2, p. 20, 2022.

[126] R. Antwi, J. D. Gadze, E. T. Tchao, A. Sikora, H. Nunoo-Mensah, A. S. Agbemenu, K. O.-B. Obour Agyekum, J. O. Agyemang, D. Welte, and E. Keelson, "A survey on network optimization techniques for blockchain systems," *Algorithms*, vol. 15, no. 6, p. 193, 2022.

[127] T. Alladi, V. Chamola, N. Sahu, V. Venkatesh, A. Goyal, and M. Guizani, "A comprehensive survey on the applications of blockchain for securing vehicular networks," *IEEE Communications Surveys & Tutorials*, 2022.

[128] A. Abdelmaboud, A. I. A. Ahmed, M. Abaker, T. A. E. Eisa, H. Albasheer, S. A. Ghorashi, and F. K. Karim, "Blockchain for iot applications: Taxonomy, platforms, recent advances, challenges and future research directions," *Electronics*, vol. 11, no. 4, p. 630, 2022.

[129] S. Venkatraman and S. Parvin, "Developing an iot identity management system using blockchain," *Systems*, vol. 10, no. 2, p. 39, 2022.

[130] K. T. Nguyen, N. Oualha, and M. Laurent, "Securely outsourcing the ciphertext-policy attribute-based encryption," *World Wide Web*, vol. 21, no. 1, pp. 169–183, 2018.

[131] S. Y. A. Zaidi, M. A. Shah, H. A. Khattak, C. Maple, H. T. Rauf, A. M. El-Sherbeeny, and M. A. El-Meligy, "An attribute-based access control for iot using blockchain and smart contracts," *Sustainability*, vol. 13, no. 19, p. 10556, 2021.

[132] E. Barka, S. S. Mathew, and Y. Atif, "Securing the web of things with role-based access control," in *Codes, Cryptology, and Information Security* (S. El Hajji, A. Nitaj, C. Carlet, and E. M. Souidi, eds.), (Cham), pp. 14–26, Springer International Publishing, 2015.

[133] A. Kesarwani and P. M. Khilar, "Development of trust based access control models using fuzzy logic in cloud computing," *Journal of King Saud University-Computer and Information Sciences*, vol. 34, no. 5, pp. 1958–1967, 2022.

[134] J. L. Herrera, H.-Y. Chen, J. Berrocal, J. M. Murillo, and C. Julien, "Context-aware privacy-preserving access control for mobile computing," *Pervasive and Mobile Computing*, vol. 87, p. 101725, 2022.

[135] B. Annane, A. Alti, L. Laouamer, and H. Reffad, "Cx-cp-abe: Context-aware attribute-based access control schema and blockchain technology to ensure scalable and efficient health data privacy," *Security and Privacy*, vol. 5, no. 5, p. e249, 2022.

[136] W. Zhang, S. Liu, and Z. Xia, "A distributed privacy-preserving data aggregation scheme for smart grid with fine-grained access control," *Journal of Information Security and Applications*, vol. 66, p. 103118, 2022.

[137] K. K. Karmakar, V. Varadharajan, P. Speirs, M. Hitchens, and A. Robertson, "Sdpm: A secure smart device provisioning and monitoring service architecture for smart network infrastructure," *IEEE Internet of Things Journal*, vol. 9, no. 24, pp. 25037–25051, 2022.

[138] M. Wazid, J. Singh, A. K. Das, S. Shetty, M. K. Khan, and J. J. Rodrigues, "Ascp-iomt: Ai-enabled lightweight secure communication protocol for internet of medical things," *IEEE Access*, vol. 10, pp. 57990–58004, 2022.

[139] H. Liu, D. Han, and D. Li, "Fabric-iot: A blockchain-based access control system in iot," *IEEE Access*, vol. 8, pp. 18207–18218, 2020.

[140] J. Wang, J. Zhu, M. Zhang, I. Alam, and S. Biswas, "Function virtualization can play a great role in blockchain consensus," *IEEE Access*, vol. 10, pp. 59862–59877, 2022.

[141] A. A. Selcuk, E. Uzun, and M. R. Pariente, "A reputation-based trust management system for p2p networks," in *IEEE International Symposium on Cluster Computing and the Grid, 2004. CCGrid 2004.*, pp. 251–258, IEEE, 2004.

[142] H. Tran, M. Hitchens, V. Varadharajan, and P. Watters, "A trust based access control framework for p2p file-sharing systems," in *Proceedings of the 38th Annual Hawaii International Conference on System Sciences*, pp. 302c–302c, IEEE, 2005.

[143] N. Saxena, G. Tsudik, and J. H. Yi, "Admission control in peer-to-peer: design and performance evaluation," in *Proceedings of the 1st ACM workshop on Security of ad hoc and sensor networks*, pp. 104–113, 2003.

[144] N. Saxena, G. Tsudik, and J. H. Yi, "Identity-based access control for ad hoc groups," in *International Conference on Information Security and Cryptology*, pp. 362–379, Springer, 2004.

[145] R. Sandhu and X. Zhang, "Peer-to-peer access control architecture using trusted computing technology," in *Proceedings of the tenth ACM symposium on Access control models and technologies*, pp. 147–158, 2005.

[146] P. Fenkam, S. Dustdar, E. Kirda, G. Reif, and H. Gall, "Towards an access control system for mobile peer-to-peer collaborative environments," in *Proceedings. Eleventh IEEE International Workshops on Enabling Technologies: Infrastructure for Collaborative Enterprises*, pp. 95–100, IEEE, 2002.

[147] N. Fotiou, I. Pittaras, V. A. Siris, S. Voulgaris, and G. C. Polyzos, "Secure iot access at scale using blockchains and smart contracts," in *2019 IEEE 20th International Symposium on" A World of Wireless, Mobile and Multimedia Networks"(WoWMoM)*, pp. 1–6, IEEE, 2019.

[148] A. Z. Ourad, B. Belgacem, and K. Salah, "Using blockchain for iot access control and authentication management," in *International Conference on Internet of Things*, pp. 150–164, Springer, 2018.

[149] R. Xu, Y. Chen, E. Blasch, and G. Chen, "Blendcac: A blockchain-enabled decentralized capability-based access control for iots," in *2018 IEEE International Conference on Internet of Things (iThings) and IEEE Green Computing and Communications (GreenCom) and IEEE Cyber, Physical and Social Computing (CPSCom) and IEEE Smart Data (SmartData)*, pp. 1027–1034, IEEE, 2018.

[150] J. Lu, R. Li, Z. Lu, and X. Ma, "A role-based access control architecture for p2p file-sharing systems using primary/backup strategy," in *2009 International Conference on Networks Security, Wireless Communications and Trusted Computing*, vol. 1, pp. 700–703, IEEE, 2009.

[151] M. S. Reza, S. Biswas, A. Alghamdi, M. Alrizq, A. K. Bairagi, and M. Masud, "Acc: Blockchain based trusted management of academic credentials," in *2021 IEEE International Symposium on Smart Electronic Systems (iSES)*, pp. 438–443, 2021.

[152] F.-K. Tseng, J. K. Zao, Y.-H. Liu, and F.-P. Kuo, "Halo: A hierarchical identity-based public key infrastructure for peer-to-peer opportunistic collaboration," in *2009 Tenth International Conference on Mobile Data Management: Systems, Services and Middleware*, pp. 672–679, IEEE, 2009.

[153] K. A. Aravind, B. R. Naik, and C. S. Chennarao, "Combined digital signature with sha hashing technique-based secure system: An application of blockchain using iot," *Turkish Journal of Computer and Mathematics Education (TURCOMAT)*, vol. 13, no. 03, pp. 402–418, 2022.

[154] A. Mohsenzadeh, A. J. Bidgoly, and Y. Farjami, "A novel reputation-based consensus framework (rcf) in distributed ledger technology," *Computer Communications*, vol. 190, pp. 126–144, 2022.

[155] K. Mrazek, B. Holton, C. Cathcart, J. Speirer, J. Do, and T. K. Mohd, "Risks in blockchain–a survey about recent attacks with mitigation methods and solutions for overall," in *2022 IEEE International Conference on Electro Information Technology (eIT)*, pp. 5–10, IEEE, 2022.

[156] S. Ramos, F. Pianese, T. Leach, and E. Oliveras, "A great disturbance in the crypto: Understanding cryptocurrency returns under attacks," *Blockchain: Research and Applications*, vol. 2, no. 3, p. 100021, 2021.

[157] I. Makarov and A. Schoar, "Cryptocurrencies and decentralized finance (defi)," tech. rep., National Bureau of Economic Research, 2022.

[158] A. Averin and O. Averina, "Review of blockchain technology vulnerabilities and blockchain-system attacks," in *2019 International Multi-Conference on Industrial Engineering and Modern Technologies (FarEastCon)*, pp. 1–6, IEEE, 2019.

[159] J. J. Kearney and C. A. Perez-Delgado, "Vulnerability of blockchain technologies to quantum attacks," *Array*, vol. 10, p. 100065, 2021.

[160] H. E. Arslanian *et al.*, "The book of crypto," *Springer Books*, 2022.

[161] Y. Zhang, M. Zhao, T. Li, Y. Wang, and T. Liang, "Achieving optimal rewards in cryptocurrency stubborn mining with state transition analysis," *Information Sciences*, 2023.

[162] M. J. Mihaljević, L. Wang, S. Xu, and M. Todorović, "An approach for blockchain pool mining employing the consensus protocol robust against block withholding and selfish mining attacks," *Symmetry*, vol. 14, no. 8, p. 1711, 2022.

[163] X. Dong, F. Wu, A. Faree, D. Guo, Y. Shen, and J. Ma, "Selfholding: A combined attack model using selfish mining with block withholding attack," *Computers & Security*, vol. 87, p. 101584, 2019.

[164] K. Nicolas, Y. Wang, G. C. Giakos, B. Wei, and H. Shen, "Blockchain system defensive overview for double-spend and selfish mining attacks: A systematic approach," *IEEE Access*, vol. 9, pp. 3838–3857, 2020.

[165] M. Kedziora, P. Kozłowski, M. Szczepanik, and P. Jóźwiak, "Analysis of blockchain selfish mining attacks," in *International Conference on Information Systems Architecture and Technology*, pp. 231–240, Springer, 2019.

[166] Q. Bai, X. Zhou, X. Wang, Y. Xu, X. Wang, and Q. Kong, "A deep dive into blockchain selfish mining," in *ICC 2019-2019 IEEE International Conference on Communications (ICC)*, pp. 1–6, IEEE, 2019.

[167] S. K. Dwivedi, R. Amin, and S. Vollala, "Smart contract and ipfs-based trustworthy secure data storage and device authentication scheme in fog computing environment," *Peer-to-Peer Networking and Applications*, pp. 1–21, 2022.

[168] Z. Wang, Z. Zheng, W. Jiang, and S. Tang, "Blockchain-enabled data sharing in supply chains: Model, operationalization, and tutorial," *Production and Operations Management*, vol. 30, no. 7, pp. 1965–1985, 2021.

[169] J. Scharfman, "Decentralized finance (defi) fraud and hacks: Part 2," in *The Cryptocurrency and Digital Asset Fraud Casebook*, pp. 97–110, Springer, 2023.

[170] N. Kshetri and J. Voas, "Blockchain's carbon and environmental footprints," *Computer*, vol. 55, no. 8, pp. 89–94, 2022.

[171] X. Yin, Y. Zhu, and J. Hu, "A comprehensive survey of privacy-preserving federated learning: A taxonomy, review, and future directions," *ACM Computing Surveys (CSUR)*, vol. 54, no. 6, pp. 1–36, 2021.

[172] K. M. Khan, J. Arshad, and M. M. Khan, "Empirical analysis of transaction malleability within blockchain-based e-voting," *Computers & Security*, vol. 100, p. 102081, 2021.

[173] R. R. Vokerla, B. Shanmugam, S. Azam, A. Karim, F. De Boer, M. Jonkman, and F. Faisal, "An overview of blockchain applications and attacks," in *2019 international conference on vision towards emerging trends in communication and networking (ViTECoN)*, pp. 1–6, IEEE, 2019.

[174] J. Stempel, "Dunkin' donuts parent settles new york cyberattack lawsuit, is fined," 2022.

[175] P. Reedy, "Interpol review of digital evidence 2016-2019," *Forensic Science International: Synergy*, vol. 2, pp. 489–520, 2020.

[176] N. Al-Otaiby, A. Alhindi, and H. Kurdi, "Anttrust: An ant-inspired trust management system for peer-to-peer networks," *Sensors*, vol. 22, no. 2, p. 533, 2022.

[177] N. C. Velayudhan, A. Anitha, and M. Madanan, "Sybil attack with rsu detection and location privacy in urban vanets: An efficient eporp technique," *Wireless Personal Communications*, vol. 122, no. 4, pp. 3573–3601, 2022.

[178] B. Hammi, Y. M. Idir, S. Zeadally, R. Khatoun, and J. Nebhen, "Is it really easy to detect sybil attacks in c-its environments: a position paper," *IEEE Transactions on Intelligent Transportation Systems*, vol. 23, no. 10, pp. 18273–18287, 2022.

[179] C. Cimpanu, "A mysterious group has hijacked tor exit nodes to perform ssl stripping attacks," 2020.

[180] nusenu, "Is "kax17" performing de-anonymization attacks against tor users?," 2021.

[181] H. Djuitcheu, M. Debes, M. Aumüller, and J. Seitz, "Recent review of distributed denial of service attacks in the internet of things," in *2022 5th Conference on Cloud and Internet of Things (CIoT)*, pp. 32–39, IEEE, 2022.

[182] T. Sunitha, V. Vijayashanthi, M. Navaneethakrishan, T. Mohanaprakash, S. Ashwin, T. Harish, and E. A. Stanes, "Key observation to prevent ip spoofing in ddos attack on cloud environment," in *Soft Computing: Theories and Applications: Proceedings of SoCTA 2022*, pp. 493–505, Springer, 2023.

[183] R. Chaganti, B. Bhushan, and V. Ravi, "The role of blockchain in ddos attacks mitigation: techniques, open challenges and future directions," *arXiv preprint arXiv:2202.03617*, 2022.

[184] A. Scroxton, "New zealand activates security services as ddos outage enters fourth day," 2020.

[185] A. Bhardwaj, V. Mangat, and R. Vig, "Hyperband tuned deep neural network with well posed stacked sparse autoencoder for detection of ddos attacks in cloud," *IEEE Access*, vol. 8, pp. 181916–181929, 2020.

[186] S. K. Verma, A. Verma, and A. C. Pandey, "Addressing dao insider attacks in ipv6-based low-power and lossy networks," in *2022 IEEE Region 10 Symposium (TENSYMP)*, pp. 1–6, IEEE, 2022.

[187] J. Xu and Y. Feng, "Reap the harvest on blockchain: A survey of yield farming protocols," *IEEE Transactions on Network and Service Management*, 2022.

[188] J. Chen, X. Xia, D. Lo, J. Grundy, and X. Yang, "Maintenance-related concerns for post-deployed ethereum smart contract development: issues, techniques, and future challenges," *Empirical Software Engineering*, vol. 26, no. 6, p. 117, 2021.

[189] D. Prashar *et al.*, "Analysis on blockchain vulnerabilities & attacks on wallet," in *2021 3rd International Conference on Advances in Computing, Communication Control and Networking (ICAC3N)*, pp. 1515–1521, IEEE, 2021.

[190] P. Dimitrios, *Distributed Consensus Inference and Blockchain*. PhD thesis, ARISTOTLE UNIVERSITY OF THESSALONIKI, 2022.

[191] E. Chang, P. Darcy, K.-K. R. Choo, and N.-A. Le-Khac, "Forensic artefact discovery and attribution from android cryptocurrency wallet applications," *arXiv preprint arXiv:2205.14611*, 2022.

[192] W. Nam and H. Kil, "Formal verification of blockchain smart contracts via atl model checking," *IEEE Access*, vol. 10, pp. 8151–8162, 2022.

[193] S. Biswas, K. Sharif, F. Li, S. Maharjan, S. P. Mohanty, and Y. Wang, "Pobt: A lightweight consensus algorithm for scalable iot business blockchain," *IEEE Internet of Things Journal*, vol. 7, no. 3, pp. 2343–2355, 2020.

[194] F. A. Aponte-Novoa, A. L. S. Orozco, R. Villanueva-Polanco, and P. Wightman, "The 51% attack on blockchains: A mining behavior study," *IEEE Access*, vol. 9, pp. 140549–140564, 2021.

[195] P. Tekchandani, I. Pradhan, A. K. Das, N. Kumar, and Y. Park, "Blockchain-enabled secure big data analytics for internet of things smart applications," *IEEE Internet of Things Journal*, vol. 10, no. 7, pp. 6428–6443, 2023.